\documentclass[aps,twocolumn,superscriptaddress,preprintnumbers,pre]{revtex4}
\usepackage{xr-hyper}
\usepackage{mathrsfs, hyperref}
\usepackage{amssymb, amsbsy, amsmath, latexsym, dsfont, array, layout, graphics,mathrsfs,braket,amsfonts,amsthm,
  amssymb,graphicx,subfigure, youngtab,color,bm,mathtools,braket,verbatim,url,cleveref,natbib,hypernat,extarrows,inputenc,multirow,bbm,dsfont}
  
\usepackage{xcolor}
\colorlet{RED}{red}
\usepackage{graphicx,psfrag}
\usepackage{amsfonts}
\usepackage[figuresright]{rotating}  
\usepackage{amssymb}
\usepackage{amsmath}
\usepackage{subfigure}
\usepackage{multirow}
\usepackage{tabularx}
\usepackage[resetlabels]{multibib}
\usepackage{amssymb, amsbsy, amsmath, latexsym, dsfont, array, layout, graphics,mathrsfs,braket,amsfonts,amsthm,
  amssymb,graphicx,subfigure,dcolumn, youngtab,color,bm,mathtools,braket,verbatim,url,cleveref,natbib,hypernat,extarrows,inputenc,multirow}
\usepackage{xcolor}
\graphicspath{ {./images/} }

\usepackage{mathrsfs, hyperref}
\usepackage{amssymb, amsbsy, amsmath, latexsym, dsfont, array, layout, graphics,mathrsfs,braket,amsfonts,amsthm,
  amssymb,graphicx,subfigure, youngtab,color,bm,mathtools,braket,verbatim,url,cleveref,natbib,hypernat,extarrows,inputenc,multirow,bbm}
\usepackage{graphicx}
\usepackage{dcolumn}
\usepackage{bm}
\usepackage{xcolor}
\usepackage[normalem]{ulem}
\usepackage[T1]{fontenc}
\graphicspath{ {./images/} }
\allowdisplaybreaks 

\newcommand{\splitatcommas}[1]{%
\begingroup 
\begingroup\lccode`~=`, \lowercase{\endgroup 
\edef~{\mathchar\the\mathcode`, \penalty0 \noexpand\hspace{0pt plus 1em}}%
}\mathcode`,="8000 #1%
\endgroup 
} 

\begin{document}

\begin{abstract}
  We study topological flat bands with distinct features that deviate from conventional Landau level behavior. We show that even in the ideal quantum geometry limit, moiré flat band systems can exhibit physical phenomena fundamentally different from Landau levels without lattices. In particular, we find new fractional quantum Hall states emerging from multi-band vortexable systems, where multiple exactly flat bands appear at the Fermi energy. While the set of bands as a whole exhibits ideal quantum geometry, individual bands separately lose vortexability, and thus making them very different from a stack of Landau levels. At certain filling fractions, we find fractional states whose Hall conductivity deviates from the filling factor. Through careful numerical and analytical studies, we rule out all known mechanisms—such as fractional quantum Hall crystals or separate filling of trivial and topological bands—as possible explanations. Leveraging the exact solvability of vortexable systems, we use analytic Bloch wavefunctions to uncover the origin of these new fractional states, which arises from the commensurability between the moiré unit cell and the magnetic unit cell of an emergent effective magnetic field. 
\end{abstract}
\title{Unconventional Fractional Phases in Multi-Band Vortexable Systems}
\author{Siddhartha Sarkar}\thanks{These two authors contributed equally}
\affiliation{%
Department of Physics, University of Michigan, Ann Arbor, MI 48109, USA
}
\affiliation{Max Planck Institute for the Physics of Complex Systems, N\"othnitzer Stra\ss e 38, 01187 Dresden, Germany}
\author{Xiaohan Wan}\thanks{These two authors contributed equally}
\affiliation{%
Department of Physics, University of Michigan, Ann Arbor, MI 48109, USA
}
\author{Ang-Kun Wu}
\affiliation{%
Theoretical Division, T-4, Los Alamos National Laboratory, Los Alamos, New Mexico 87545, USA}
\author{Shi-Zeng Lin}
\email{szl@lanl.gov}
\affiliation{%
Theoretical Division, T-4 and CNLS, Los Alamos National Laboratory, Los Alamos, New Mexico 87545, USA
}
\affiliation{%
Center for Integrated Nanotechnologies (CINT), Los Alamos National Laboratory, Los Alamos, New Mexico 87545, USA
}
\author{Kai Sun}
\email{sunkai@umich.edu}
\affiliation{%
Department of Physics, University of Michigan, Ann Arbor, MI 48109, USA
}
\maketitle
\noindent\textit{Introduction.}--
The interplay between band topology and strong interactions in Landau levels (LLs) gives rise to exotic quantum states, known as the fractional quantum Hall (FQH) effect~\cite{stormer1999fractional,laughlin1999nobel}. Inspired by Haldane’s seminal work on the quantum anomalous Hall effect~\cite{haldane1988model}, it was predicted that fractionalized states could also arise without magnetic fields, by utilizing topological flat bands—leading to the concepts of fractional quantum anomalous Hall effects and fractional Chern insulators (FCIs)\cite{tang2011high, sun2011nearly, neupert2011fractional, sheng2011fractional, regnault2011fractional,bernevig2012emergent}. Recent proposals suggested moiré systems as ideal platforms\cite{repellin2020chern,ledwith2020fractional,liu2021gate,li2021spontaneous,wu2024quantum}, and FCI states have now been observed in twisted bilayer MoTe$_2$\cite{cai2023signatures,zeng2023thermodynamic,park2023observation,xu2023observation} and rhombohedral graphene/hBN superlattices\cite{lu2024fractional,xie2024tunable}.

Although FCIs do not require external magnetic fields, most known examples are adiabatically connected to FQH states in LLs or multilayered LL systems~\cite{regnault2011fractional,bernevig2012emergent,wu2012,wu2012zoology,sterdyniak2013series,wu2015fractional}. It has also been shown that topological flat bands with quantum geometry more analogous to Landau levels—namely, with more uniform Berry curvature distribution and ideal quantum geometry (meaning that the trace of the quantum metric tensor, which measures the spread of the wavefunctions~\cite{vanderbilt2018berry}, equals the Berry curvature)—host more robust fractional states with larger energy gaps~\cite{wu2012,parameswaran2013fractional,roy2014band,wu2015fractional,ledwith2020fractional,ledwith2021strong,wang2021exact,mera2021engineering,ledwith2023vortexability}. In this context, the concept of ``vortexable'' bands~\cite{ledwith2023vortexability} have recently been put forward; these bands are defined as bands with wavefunctions $\psi$ such that after multiplying the wavefunction by a holomorphic function $f(z)$ of complexified real space coordinate $z=x+iy$, the new function $f(z)\psi$ is fully contained in the band.
These bands share the same exotic properties as the remarkable lowest Landau level (LLL)—such as ideal quantum geometry and nontrivial algebra—and exhibit corresponding experimental signatures: for example, vortex insertion costs no energy, fractional states can emerge from short-range repulsion~\cite{ledwith2020fractional,ledwith2021strong,wang2021exact,mera2021engineering,ledwith2022family,ledwith2023vortexability}.

In this study, we explore the opposite question: can topological flat bands support fractionalized physics that fundamentally deviates from LL behavior? To address this, we focus on exactly flat bands with ideal quantum geometry as a controlled platform, while emphasizing that our conclusions extend beyond the ideal limit due to many-body gap protection. Recent studies~\cite{tarnopolsky2019origin,ledwith2022family,wang2022hierarchy,wan2023topological,eugenio2022twisted,le2022double,becker2022fine,becker2023degenerate,sarkar2023symmetry} reveal that vortexable systems encompass a rich family of models: while some closely resemble LLs, others exhibit fundamentally new behaviors. These systems preserve advantages such as flatness and exact solvability, yet display features absent in conventional LLs.

A striking example is the emergence of multi-band vortexable systems~\cite{wan2023topological,le2022double,becker2022fine,becker2023degenerate,sarkar2023symmetry}, where multiple exactly flat bands appear at the Fermi energy. As a whole, they remain vortexable with ideal quantum geometry, but individually lose vortexability—unlike stacks of LLs, where each band retains this property. Moreover, instead of carrying total Chern number $C=N$ as in $N$ LLs, these systems universally exhibit $C=1$, independent of degeneracy or symmetry~\cite{sarkar2023symmetry}.
\begin{figure}
     \centering
\includegraphics[scale=0.9]{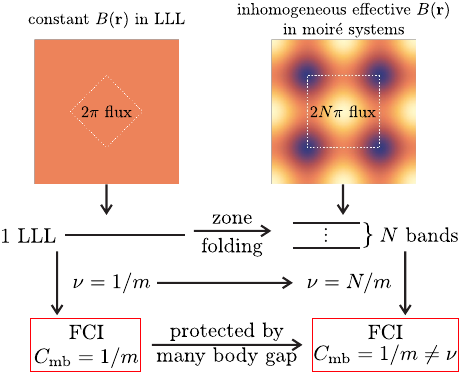}
     \caption{Schematic showing how an FCI state with unequal filling fraction and many-body Chern number arises in multiband vortexable systems due to inhomogeneous effective magnetic field distribution $B(\mathbf{r})$ and commensurability between moir\'e unit cell and magnetic length. The white dashed lines show the smallest unit cells.}
     \label{fig:0}
\end{figure}

Using exact numerics, we identify fractional states across various multi-band vortexable models. Depending on filling, the behavior varies: some mimic $N-1$ trivial bands plus a $C=1$ Chern band, while others exhibit fundamentally new fractionalization. For instance, at $\nu=2/3$, we observe a fractional state with many-body Chern number $C_\text{mb}=1/3$, violating the conventional relation $C_\text{mb}=\nu$. Through detailed numerical and theoretical analyses, we rule out known explanations such as fractional crystals or simple band fillings. Crucially, the many-body ground states show strong inter-band entanglement, defying descriptions based on individual bands.

Leveraging the exact solvability of vortexable systems, we analytically solve for the Bloch wavefunctions and identify the underlying mechanism: the commensurability between the moiré unit cell and the effective magnetic unit cell of an emergent magnetic field. In contrast to LLs (without lattices), where a single magnetic length governs physics, topological flat bands feature two length scales—the moiré unit cell and an emergent magnetic length. Spatial periodicity requires that the unit cell encloses an integer ($N$) number of flux quanta. When $N=1$, the system mimics a LL; when $N>1$, new types of fractional states emerge, as schematically shown in Fig.~\ref{fig:0}. Our study focuses on this $N>1$ regime, revealing novel fractionalization fundamentally distinct from conventional Landau levels.

\begin{figure}[t]
     \centering
\includegraphics[scale=0.95]{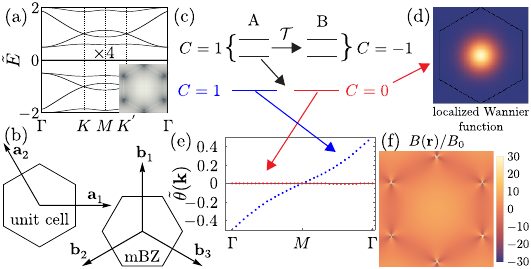}
     \caption{Four exact flat bands in single layer system with QBCP under periodic strain field. 
     (a) Band structure for the Hamiltonian in Eq.~\eqref{eq:hamiltonian} with $\tilde{A}(\mathbf{r})= -\frac{\alpha}{2}\sum_{n=1}^{3}e^{i(4-n)\phi} \cos\left(\mathbf{b}_n \cdot \mathbf{r} \right)$ showing 4 exact flat bands at the ``magic'' value  $\tilde{\alpha} =\frac{\alpha}{|\mathbf{b}| ^{2}} = -2.88$ where $\mathbf{b}_i (i=1, 2, 3)$ are the moir\'e reciprocal lattice vectors shown in (b). (b) also shows the moir\'e unit cell, lattice vectors $\mathbf{a}_1$ and $\mathbf{a}_2$, moir\'e Brillouin zone (mBZ). Density plot of $|\psi_{\Gamma}(\mathbf{r})|$ (normalized by its maximum) is shown at the bottom right corner of (a). The dark spots represent zeros of $|\psi_{\Gamma}(\mathbf{r})|$. There are two zeros at the corners of the unit cell.
     (c) Decomposition of the four flat bands in sublattice space. Two bands which are polarized on A sublattice together have $C=1$, while the other two bands polarized on B sublattice together have $C=-1$. The two sublattices are related by time reversal symmetry $\mathcal{T}$. The two bands polarized on A sublattice can be decomposed into one $C=1$ band and one $C=0$ band. The $C=0$ band corresponds to a very localized Wannier function $w(\mathbf{r})$. (d) Density plot of $|w(\mathbf{r})|^2$, where brighter color represent larger value of the norm of the wavefunction.
     (e) Wilson loop spectra $\tilde{\theta}({\mathbf{k}})=\frac{\theta({\mathbf{k}})}{2\pi}$ of the $C=1$ and $C=0$ band plotted in blue and red, respectively.
     (f) Effective magnetic field $B(\mathbf{r})/B_0$ in the moir\'e unit cell for the flatbands  obtained by comparing the flatband wavefunctions to lowest Landau Level (LLL) wavefunctions.
    }
     \label{fig:1}
\end{figure}

\noindent\textit{Model.}--Although our results are generic to any continuum model with multiple flat bands per sublattice/spin/valley with ideal quantum geometry, as a concrete example, we consider a continuum Hamiltonian which descibes a quadratic band crossing point (QBCP) at a time reversal invariant momentum coupled to a moir\'e periodic strain field~\cite{wan2023topological},
\begin{equation}\label{eq:hamiltonian}
    \mathcal{H}(\mathbf{r})
    = \begin{pmatrix}0 &\text{h.c.}\\ -4 \overline{\partial_{z}}^2 +\tilde{A}(\mathbf{r}) & 0\end{pmatrix}
\end{equation}
Here $z=x+iy$ is the complex coordinate, $\tilde{A}(\mathbf{r}) = A_x(\mathbf{r})+iA_y(\mathbf{r})$ where $A_x=u_{xx}-u_{yy}$ and $A_y=u_{xy}$ are moir\'e periodic shear strain fields. Note that this Hamiltonian is chiral/sublattice symmetric $\{\sigma_z,\mathcal{H}(\mathbf{r})\}=0$. As shown in Fig.~\ref{fig:1}(a), it was shown in~\cite{wan2023topological} that for a simple first harmonic moir\'e strain field with $p6mm$ space group symmetry there are four exact flat bands at some magic value of strain strength. Two of the four bands are polarized on the A sublattice while the other two are polarized on B sublattice. Notably, one of the two sublattice-polarized $\Gamma$ point wavefunctions $\psi_\Gamma(\mathbf{r})$ has two zeros (as shown in  Fig.~\ref{fig:1}(a)) at the corners $\mathbf{r}_0^{(1)}$ and $\mathbf{r}_0^{(2)}$ of the unit cell. Using these two zeros two holomorphic Bloch-periodic functions $f_\mathbf{k}^{(1)}(z)$ and $f_\mathbf{k}^{(2)}(z)$ can be constructed (Supplemental Material (SM)~\cite{SM2022})) such that the two Bloch-periodic wave-functions $\psi_{\mathbf{k}}^{(i)}(\mathbf{r}) = f_\mathbf{k}^{(i)}(z)\psi_\Gamma(\mathbf{r})$ $(i=1,2)$ are zero modes of the Hamiltonian, and hence they are the exact wave-functions of the two sublattice-polarized flatbands. 
The two flat bands polarized to one sublattice together have Chern number $C=1$. 
Furthermore, due to the form of the Hamiltonian, for any flatband wavefunction $\psi_\mathbf{k}^{(i)}(\mathbf{r})$ defined above, $f(z)\psi_\mathbf{k}^{(i)}(\mathbf{r})$ is also a zero mode for any holomorphic function $f(z)$ of $z$, and hence contained in the two flatbands; this implies that these bands are vortexable~\cite{ledwith2023vortexability}.
All of these properties mentioned until now are generic to all examples of multiple sublattice-polarized flatbands found in different models~\cite{le2022double,becker2022fine,becker2023degenerate,sarkar2023symmetry}.
Importantly, the two sublattice-polarized bands can be decomposed (we perform this decomposition using the software Wannier90~\cite{marzari1997maximally,souza2001maximally,pizzi2020wannier90}, see SM~\cite{SM2022}) into a topologically trivial band with very localized Wannier function and a topological band with Chern number $C=1$ (see Figs.~\ref{fig:1}(c-e)). After this decomposition, the Chern band has ideal quantum geomtry, but the trivial band does not have ideal quantum geometry as we show in SM~\cite{SM2022}. However, it is worth reiterating that the two bands together {\color{black} are vortexable}~\cite{sarkar2023symmetry}.
To study interacting states in these four flat bands, we note that if the energy scale of interaction is much smaller than the single particle gap between the flatbands and remote bands, we can project the interacting Hamiltonian onto the flatbands. When filling fraction is $\nu=2$, 
exchange interaction drives all electrons to the two bands polarized on one of the sublattices (Hund's rule), and the resulting ground state is a spontaneous time reversal symmetry breaking Chern insulator. We assume that for $\nu<2$, electrons are still polarized on one of the sublattices, and write down the two-band (polarized on the same sublattice) projected interacting Hamiltonian
\begin{equation}
     H_\text{int} = \frac{1}{2A} \sum_\mathbf{q} V(\mathbf{q})\rho(\mathbf{q})\rho(-\mathbf{q}),
\end{equation}
where $A$ is the system area, $V(\mathbf{q})/A = 4\pi U \tanh(dq)/(N_s\sqrt{3}qa)$ is the screened Coulomb interaction, $d$ is the separation between the two electrodes, $a$ is the moir\'e lattice constant, $q=|\mathbf{q}|$, $N_s$ is the number of moir\'e unit cells, $U$ is the bare Coulomb energy between two particles at a distance $a$, and we introduced the projected density operator $\rho(\mathbf{q}) = \sum_{\mathbf{k}}\lambda_{i,j,\mathbf{q}}(\mathbf{k})c^\dagger_{\mathbf{k},i}c_{\mathbf{k}+\mathbf{q},j}$, where $\lambda_{i,j,\mathbf{q}}(\mathbf{k}) = \langle \psi_\mathbf{k}^{(i)}(\mathbf{r})|e^{-i\mathbf{q}\cdot\mathbf{r}}|\psi_{\mathbf{k}+\mathbf{q}}^{(j)}(\mathbf{r})\rangle$ is the form factor, and $i,j$ run over the Chern ($C$) and Wannier ($W$) bands. Note that $\rho(\mathbf{q})$ can be split into three parts $\rho(\mathbf{q}) = \rho_{CC}(\mathbf{q})+\rho_{WW}(\mathbf{q})+\rho_{CW}(\mathbf{q})$, where the first two terms are the densties in the Chern band and the trivial band, respectively, and the third term is a tunneling term between the two bands (this term breaks the $U(1)\times U(1)$ symmetry). Below we present exact diagonalization results at different filling fractions.

\begin{figure}[t]
     \centering
\includegraphics[scale=0.85]{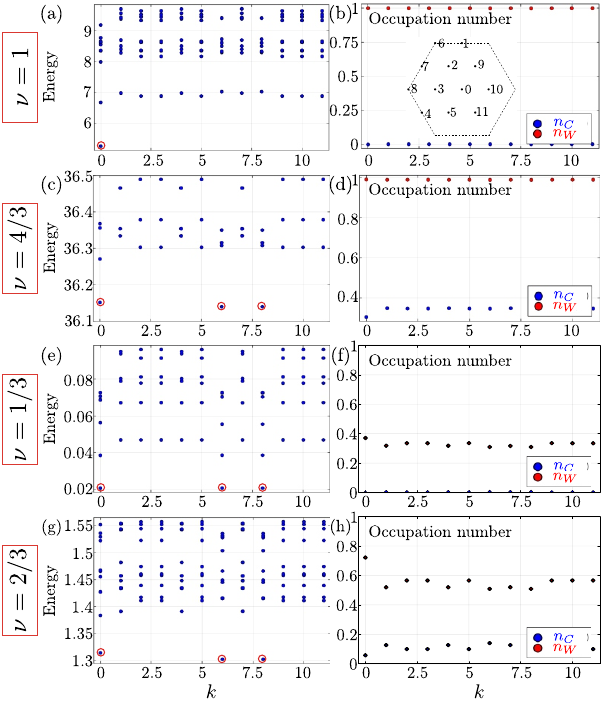}
     \caption{Many-body energy spectra (a,c,e,g) and occupation number in the Chern ($n_C(\mathbf{k}) = \langle c^\dag_C(\mathbf{k})c_C(\mathbf{k})\rangle$) and Wannier ($n_W(\mathbf{k}) = \langle c^\dag_W(\mathbf{k})c_W(\mathbf{k})\rangle$) bands of the ground-states (averaged over the three ground-states) (b,d,f,h) at filling $\nu=1$, $4/3$, $1/3$ and $2/3$, respectively. The momentum numbers are shown in the inset in (b). The ground-states in (a,c,e,g) are marked by red circles. At $\nu=1$, there is a single insulating ground-state, whereas for each of the other filling fractions, there are three quasi-degnerate ground states. The many-body Chern numbers of the ground states are  $C_\text{mb}=0$ in (a) and (e), and  are $C_\text{mb}=1/3$ for (c) and (g). For $\nu=1$ and $\nu=1/3$ ground states, (b) and (f) show that the Chern band is fully empty, all electrons are in the Wanier band. For $\nu=4/3$ ground state, (d) shows the Wannier band is fully filled, and the Chern band is $1/3$ filled. For $\nu=2/3$ ground state, (h) shows that both bands are partially occupied.
    }
     \label{fig:2}
\end{figure}

\noindent\textit{Filling fractions $\nu = 1$, $\nu=4/3$, $\nu=5/3$ and $\nu=1/3$.}--At Filling fraction $\nu=1$, we find an insulating ground state (Fig.~\ref{fig:2}(a)) with the trivial band fully occupied and the Chern band empty as can be seen from the occupation number plot of the ground state shown in Fig.~\ref{fig:2}(b). This can be understood from the fact that the Wannier functions of the trivial band are very localized (Fig.~\ref{fig:1}(d)) meaning that the trace of the quantum metric $\text{tr}(g(\mathbf{k}))$ of this band is small (see {\color{black}SM Fig.~S2} for the plot of $\text{tr}(g(\mathbf{k}))$), whereas $\text{tr}(g(\mathbf{k}))$ of the Chern band is much larger since it is bounded by the Berry curvature. Since Fock contribution of the interaction energy is roughly proportional to $\text{tr}(g(\mathbf{k}))$~\cite{wu2024quantum}, by occupying the trivial band the electrons minimize the Fock contribution. We numerically verified that the ground state has many-body Chern number $C_\text{mb} = 0$ via flux insertion (we follow~\cite{wu2024quantum} for $C_\text{mb}$ calculation). At filling $\nu=4/3$, we find three quasi-degenerate ground states at momenta consistent with both a CDW and FCI state shown in Fig.~\ref{fig:2}(c). We numerically find that  the trivial band is fully occupied, whereas the Chern band is $1/3$ filled (Fig.~\ref{fig:2}(d)), which is consistent with our numerical finding of $C_\text{mb}=1/3$. We show in SM~\cite{SM2022} that the particle entanglement spectrum (PES)~\cite{li2008entanglement,sterdyniak2011extracting} of the ground states is also consistent with a FCI state. Filling fraction $\nu=5/3$ is similar to $\nu=4/3$, the Chern band is two-third filled in this case, giving rise to a FCI with $C_\text{mb}=2/3$ (see SM~\cite{SM2022}). At $\nu=1/3$, there are three quasi-degnerate ground states  with center of mass momenta at zone center ($\Gamma$) and zone corners ($K$ and $K'$). (Fig.~\ref{fig:2}(e)), with all electrons in the Wannier band. This is a $\sqrt{3}\times\sqrt{3}$ CDW state; we show that density-density correlation function of this ground state has pronounced peaks at the corner of the BZ (see SM~\cite{SM2022}).

\noindent\textit{Filling fraction $\nu=2/3$}.--From the discussion above, na\"ively one may assume that for $\nu\leq 1$, all electrons occupy the trivial band, and for $\nu>1$, the trivial band is fully filled and the rest of the electrons occupy the Chern band. However, remarkably, this simple picture does not hold for $\nu=2/3$. At this filling, we find three quasi-degenerate ground states shown in Fig.~\ref{fig:2}(g), which is consistent with both CDW and FCI. However, there are a few puzzling properties of these ground-states: (i) we numerically find the many-body Chern number $C_\text{mb}=1/3$, (ii) the occupation number is nonzero in both bands as shown in Fig.~\ref{fig:2}(h), (iii) we also find (see {\color{black}Appendix A}) that the off-diagonal terms in the occupation number matrix $n_{ij}(\mathbf{k}) = \langle c^\dag_i(\mathbf{k})c_j(\mathbf{k})\rangle$ ($i,j\in\{C,W\}$) as well as its determinant are nonzero at generic momenta which implies this FCI cannot be single-band-like in any basis (see {\color{black}SM Sec-XI}), (iv) in the PES of the ground states (see {\color{black}Appendix C}) the number of states below the gap is consistent with $\nu=1/3$ FCI state in a single Chern band with twice the number of single particle states as each band of this system has, (v) we show in {\color{black}SM}~\cite{SM2022} that the tunneling term $\rho_{CW}(\mathbf{q})$ in the denisty operator is crucial for this state, as all electrons occupy the trivial band and form a topologically trivial $\sqrt{3}\times\sqrt{3}$ CDW state if we artificially eliminate this term from $H_\text{int}$ to restore $U(1)\times U(1)$ symmetry. Below we explain this apparent discrepancy between the filling fraction and many-body Chern 
number.

\begin{figure}[t]
     \centering
\includegraphics[scale=0.95]{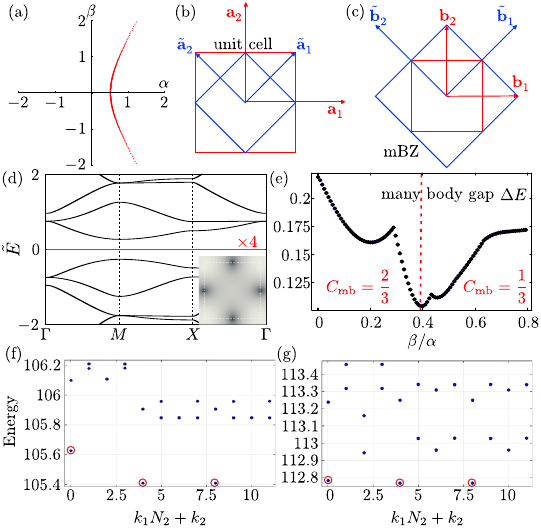}
     \caption{Phase transition between FCIs with $C_\text{mb}=2/3$ and $C_\text{mb}=1/3$ at filling fraction $\nu=4/3$ in a $p4mm$ symmetric system. The single particle Hamiltonian is given in Eqs.~\eqref{eq:hamiltonian} and~\eqref{eq:p4mmStrain}. 
     The unit cell, mBZ, (reciprocal) lattice vectors are shown in red in (b) and (c). However,
     when $\beta = 0$, the primitive unit cell and corresponding mBZ are shown in blue in (b) and (c).
     (a) Line (in red) of ``magic'' parameters $(\alpha,\beta)$, at which four exact flatbands (two per sublattice) appear. The two flatbands per sublattice have total Chern number $|C|=1$. At the point  where the red line crosses $\alpha$ axis ($\alpha=0.5279$ and $\beta=0$), the four flatbands (two per sublattice) corresponding to the red mBZ in (c) can be unfolded into two flatbands (one per sublattice) corresponding to the blue mBZ in (c).
     (d) Band structure for $\alpha=0.6651,\beta=0.7183$ containing 4 exact flatbands at $\tilde{E}=0$. At the bottom right corner, the density plot of wavefunction $|\psi_\Gamma(\mathbf{r})|$ is shown. The dark spots in the wavefunctions are the zeros.
     (f, g) Many-body energy spectra at filling $\nu=4/3$ for $\alpha=0.5279,\beta=0$ and $\alpha=0.6651,\beta=0.7183$ respectively. In both cases, we find three quasi degenerate ground states (encircled with red circles). We numerically find the many-body Chern numbers of these quasi-degenerate ground states to be $C_\text{mb}=2/3$ and $C_\text{mb}=1/3$ for (f) and (g), respectively.
     (e) Many-body gap $\Delta E$ between the lowest energy excited state and highest energy ground state (among the three quasi-degnerate ground state) as the ratio $\beta/\alpha$ is varied along red line in (a). 
     }
     \label{fig:4}
\end{figure}

\noindent\textit{The two flatbands per sublattice are equivalent to LLL with $4\pi$ flux per magnetic unit cell.}--Recall that one of the flatband wavefunctions at $\Gamma$ point  $\psi_\Gamma(\mathbf{r})$ has two zeros in the moir\'e unit cell. It is known that LLL wavefunctions on a torus have one zero per magnetic unit cell with $2\pi$ flux~\cite{haldane1985periodic}. 
This gives the first clue that the two flatbands may be one LLL band in a spatially varying magnetic field with $4\pi$ flux per unit cell, which makes the LLL band fold into two bands.
To make it quantitative, we first note that LLL wavefunctions in a spatially varying magnetic field can be written as $\psi^\text{LLL}(\mathbf{r}) = h(z)e^{-K(\mathbf{r})/4\ell_B^2}$, where $h(z)$ is a holomorphic function, and $B(\mathbf{r}) = B_0\partial_z\overline{\partial_z}K(\mathbf{r})$~\cite{ledwith2020fractional} ($B_0 = 2\pi\hbar/e A_\text{u.c.}$, where $A_\text{u.c.}$ is the area of the magnetic unit cell, and $\ell_B^2 = \hbar/eB_0$). Similarly, writing our flatband wavefunctions $\psi_\mathbf{k}^{(i)}(\mathbf{r})$ in the form $\psi_\mathbf{k}^{(i)}(\mathbf{r})=h_\mathbf{k}^{(i)}(z)e^{-K(\mathbf{r})/4\ell_B^2}\zeta(\mathbf{r})$~\cite{dong2023composite}, we extracted the effective magnetic field per unit cell $B(\mathbf{r}) = B_0\partial_z\overline{\partial_z}K(\mathbf{r})$ as shown in Fig.~\ref{fig:1}(f) (the magnetic field distribution has interesting singularities, see SM~\cite{SM2022} for details). The total flux per moir\'e unit cell of this magnetic field is indeed $4\pi$. \textit{Hence, the two flatbands together are similar to one LLL band in a spatially varying magnetic field, and the filling fraction $\nu=2/3$ is really filling fraction $\nu_\text{unfolded}=1/3$ of this LLL band, which gives in FQH state with many-body Chern number $C_\text{mb}=1/3$}. It is worth mentioning that even though the flux of per moir\'e unit cell is $4\pi$, smaller moir\'e unit cell with $2\pi$ cannot be chosen without breaking threefold rotation symmetry $\mathcal{C}_{3z}$ due to the spatial variation of the magnetic field;  however, it is possible in a $C_{4z}$ symmetric system without breaking any rotation symmetries. In Appendix B, we show explicitly in a $\mathcal{C}_{4z}$ symmetric moir\'e system that this $\nu=2/3$ but $C_\text{mb}=1/3$ state is adiabatically connected to a $\nu_\text{unfolded}=1/3$ state of a $C=1$ Chern band that is folded into a smaller mBZ. 

\noindent\textit{Competing states at filling $\nu=4/3$.}--We showed above that for the two flatbands per sublattice with total Chern number $C=1$ and ideal quantum geometry, the ground-states at different filling fractions can be understood using one of the two interpretations of the bands: (i) one Chern band, and one trivial band, (ii) one $C=1$ LLL in a spatially varying magnetic field with total $4\pi$ flux per unit cell. \textit{The fillings $\nu = 1/3, 1, 5/3$ correspond to $\nu_{\text{unfolded}} = 1/6,1/2,5/6$ having even denominator, hence in the single band interpretation there is no FCI state for these fillings, so in this interpretation the ground state would generically be metallic, which lose energetically to the insulating ground state from the first interpretation. On the other hand, the filling fractions $\nu=2/3,4/3$ correspond to $\nu_{\text{unfolded}} = 1/3,2/3$, and both interpretations lead to gapped ground states with different topological properties, and numerics is needed to know which ground state is realized.} \textit{At $\nu=4/3$, where the first interpretation predicts an FCI state with $C_\text{mb}=1/3$ as discussed earlier (see Fig.~\ref{fig:2}(c,d)), whereas within the second interpretation, we have $\nu_{\text{unfolded}} = 2/3$, which predicts a FCI state with $C_\text{mb}=2/3$.} To show a competition between these two states we take the same Hamiltonian in Eq.~\eqref{eq:hamiltonian}, but choose a $C_{4z}$ symmetric strain field
\begin{equation}\label{eq:p4mmStrain}
\begin{split}
    \tilde{A}(\mathbf{r})=& -2i \alpha\left(\cos\left(\tilde{\mathbf{b}}_1 \cdot \mathbf{r} \right)-\cos\left(\tilde{\mathbf{b}}_2 \cdot \mathbf{r} \right)\right)\\
    &+2\beta\left(\cos\left(\mathbf{b}_1 \cdot \mathbf{r} \right)-\cos\left(\mathbf{b}_2 \cdot \mathbf{r} \right)\right),
\end{split}
\end{equation}
where $\mathbf{b}_i$ and $\tilde{\mathbf{b}}_i$ are shown in Fig.~\ref{fig:4}(c). 
When $\beta = 0$, the primitive unit cell (mBZ) is half (twice) the size of the unit cell for $\beta\neq0$, as shown in Fig.~\ref{fig:4}(b,c). Now, by varying $\alpha$ and $\beta$, we find a line in the $\alpha-\beta$ plane (shown in red in Fig.~\ref{fig:4}(a)) where there are two flatbands per sublattice at $\tilde{E}=0$ as shown in Fig.~\ref{fig:4}(d) (see SM~\cite{SM2022} for details of how we find this line). For any $(\alpha,\beta)$ on the red line, the wavefunction $\psi_\Gamma(\mathbf{r})$ has two zeros in the unit cell (Fig.~\ref{fig:4}(d)). Using this two zeros two flatband wavefunctions can be constructed just like the $p6mm$ case in Fig.~\ref{fig:1}. The two flatbands per sublattice together have Chern number $|C|=1$. We show the effective magnetic field for these flatbands in the SM~\cite{SM2022}, and verify that the magnetic flux per moir\'e unit cell is $4\pi$. Hence, these flatbands are really like the flatbands in Fig.~\ref{fig:1}, just with a different symmetry. Crucially, at the point where $\beta = 0$ in Fig.~\ref{fig:4}(a), the two flatbands per sublattice in the red mBZ in Fig.~\ref{fig:4}(c) can be unfolded to one flatband with $|C|=1$ per sublattice in the blue mBZ in Fig.~\ref{fig:4}(c). The many-body spectra in Fig.~\ref{fig:4}(f) and (g) for $\nu=4/3$ show triply degenerate ground states at ($\alpha=0.5279,\beta=0.0$) and ($\alpha=0.6651,\beta=0.7183$), respectively; however, in the former case $C_\text{mb}=2/3$ and in the latter case $C_\text{mb}=1/3$. The former one can be understood from the second picture (single unfolded $|C|=1$ with $\nu_\text{unfolded} = 2/3$), whereas the latter can be understood from the first picture (one Chern band 1/3 filled, and one trivial band fully filled). The PES of these ground states agree with these interpretations as we show in {\color{black}Appendix D}. Note that along the whole red line in Fig.~\ref{fig:4}(a) the single particle gap does not close; however, as $\beta/\alpha$ is varied, we numerically find that for $\beta/\alpha<0.39$, the many-body Chern number of ground states is $C_\text{mb}=2/3$, whereas for $\beta/\alpha>0.39$, the many-body Chern number of ground states is $C_\text{mb}=1/3$. We show the many-body gap as a function of $\beta/\alpha$ in Fig.~\ref{fig:4}(e) along the red line in Fig.~\ref{fig:4}(a). The many-body gap has a global minimum around $\beta/\alpha\approx0.39$; we conjecture that the many-body gap does not fully close due to finite size effect.

\noindent\textit{Discussion.}--One interesting platform where the physics described can be realized is twisted bilayer graphene with spatially alternating magnetic field as described in~\cite{le2022double}. This system has four flat bands (two per sublattice) per valley per spin, and in the chiral limit the high symmetry momentum wavefunction $\psi_\Gamma(\mathbf{r})$ has two zeros~\cite{sarkar2023symmetry} at the corners of the moir\'e unit cell indicating that the magnetic flux per moir\'e unit cell is $4\pi$. Also, promising 2D materials with quadratic band crossing at the Fermi level such as PO~\cite{zhu2016blue}, Mg$_2$C~\cite{wang2018monolayer}, Pd$_3$P$_2$S$_8$~\cite{park2020kagome}, Cu$_2$N~\cite{hu2023realization}, metal organic framework material Cr$_3$(HAB)$_2$~\cite{feng2024quadratic} along with rapidly improving experimental techniques for applying periodic strain field~\cite{jiang2017visualizing,zhang2019magnetotransport,mao2020evidence,cho2021highly,zhang2024patternable,zhang2024enhancing,kim2023strain} means that experimental realization of the Hamiltonian in Eq.~\eqref{eq:hamiltonian} will soon be possible. Lastly, although we focus mostly on moir\'e structure, it is worth noting that the key conclusions hold more generally for other topological flat bands with similar flatness and wavefunction characteristics. 

\let\oldaddcontentsline\addcontentsline
\renewcommand{\addcontentsline}[3]{}
\begin{acknowledgments}
\noindent \textit{Acknowledgments}.--The authors thank Patrick Ledwith for pointing out the similarity between multiple flatbands in moir\'e systems and lowest Landau level with $2n\pi$ flux per unit cell. SS thanks Debanjan Chowdhury for insightful discussions. This work was supported in part by Air Force Office of Scientific Research MURI FA9550-23-1-0334 and the Office of Naval Research MURI N00014-20-1-2479 (XW, SS and KS) and Award N00014-21-1-2770 (XW and KS), and by the Gordon and Betty Moore Foundation Award N031710 (KS). The work at LANL (SZL) was carried out under the auspices of the U.S. DOE NNSA under contract No. 89233218CNA000001 through the LDRD Program, and was supported by the Center for Nonlinear Studies at LANL, and was performed, in part, at the Center for Integrated Nanotechnologies, an Office of Science User Facility operated for the U.S. DOE Office of Science, under user proposals $\#2018BU0010$ and $\#2018BU0083$.
\end{acknowledgments}

\bibliographystyle{apsrev4-1}
\bibliography{ref}
\clearpage
\appendix
\setcounter{equation}{0}  
\setcounter{figure}{0}
\renewcommand{\theequation}{A\arabic{equation}}
\renewcommand{\thefigure}{A\arabic{figure}}
\section{Appendix A. Details of occupation number matrix $n_{ij}(\mathbf{k})$ for $(\nu=2/3$, $C_\text{mb}=1/3)$ FCI ground-states in Fig.~\ref{fig:2}(g)}

\begin{figure}[h]
     \centering
\includegraphics[width=\linewidth]{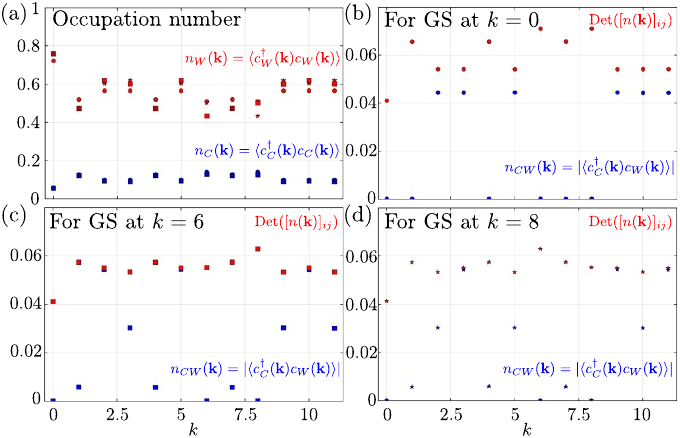}
     \caption{(a) shows the diagonal terms $n_C(\mathbf{k})=\langle c^\dag_C(\mathbf{k}) c_C(\mathbf{k})\rangle$ (in blue) and  $n_W(\mathbf{k})=\langle c^\dag_W(\mathbf{k}) c_W(\mathbf{k})\rangle$ (in red) of the occupation number matrix $n_{ij}(\mathbf{k})$ for the three quasi-degenerate ground states in Fig.~\ref{fig:2}(g). The circle, square and star shaped markers correspond to the ground states at $k=0$ (zone center), $k=6$ and $k=8$ (zone corners), respectively. (b-d) show the absolute value of off-diagonal component $n_{CW}(\mathbf{k})=|\langle c^\dag_C(\mathbf{k}) c_W(\mathbf{k})\rangle|$ (in blue) and the determinant (in red) of the occupation number matrix $n_{ij}(\mathbf{k})$ for the ground-states at momenta $k=0$, $k=6$ and $k=8$ in Fig.~\ref{fig:2}(g), respectively.
    }
     \label{fig:s8}
\end{figure}

\section{Appendix B. $(\nu=2/3, C_\text{mb}=1/3)$ state is adiabatically connected to a $\nu_\text{unfolded}=1/3$ state of a $C=1$ Chern band that is folded into a smaller mBZ.}

We noted in the main text that a $\mathcal{C}_{3z}$ symmetric moir\'e unit cell with $4\pi$ flux cannot be halved to obtain a unit cell with $2\pi$ flux without breaking $\mathcal{C}_{3z}$ symmetry. However, a $\mathcal{C}_{4z}$ symmetric moir\'e unit cell with $4\pi$ can be halved in some limits. To this end, we take the same setup as in Fig.~\ref{fig:4}(a-c).
We find three-fold quasi-degenerate ground states at filling $\nu=2/3$ everywhere along the red line (Fig.~\ref{fig:s2}(b,c)), and the many-body gap remains open on the entire line as shown in Fig.~\ref{fig:s2}(d) and the many-body Chern number remains $C_\text{mb}=1/3$, hence these ground-states along the red line are adiabatically connected.   However, FCI state at $\nu=2/3$ filling corresponding to the red mBZ for the $\beta=0$ point (see Fig.~\ref{fig:s2}(b)) is really a FCI state at $\nu_\text{unfolded}=1/3$ filling corresponding to the larger blue mBZ. Hence, the fact that it has $C_\text{mb}=1/3$ is expected. Since the FCI states at other points on the red line are adiabatically connected to this FCI at the $\beta=0$ point, their many-body Chern number must be $C_\text{mb}=1/3$ even though for these other red points the mBZ cannot be unfolded. Furthermore, in {\color{black}Appendix C}, we show that the PES of these ground states have the same counting of low-lying spectra as the $\nu=2/3$ FCI state of the $p6mm$ system in Fig.~\ref{fig:2}(g,h), supporting our claim that all these $\nu=2/3$ FCI states are the same.

\begin{figure}[h]
     \centering
\includegraphics[width=\linewidth]{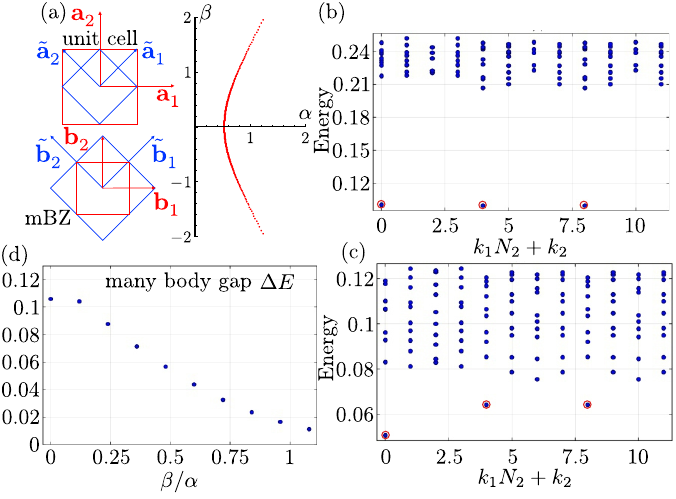}
     \caption{Evidence that the $(\nu=2/3, C_\text{mb}=1/3)$ FCI state is adiabatically connected to a $(\nu_\text{unfolded}=1/3, C_\text{mb}=1/3)$ FCI state in a $p4mm$ symmetric system. The single particle Hamiltonian is given in Eqs.~\eqref{eq:hamiltonian} and~\eqref{eq:p4mmStrain}. When $\beta = 0$ in Eq.~\eqref{eq:p4mmStrain}, the primitive unit cell and corresponding mBZ are shown in blue in (a).
     (a) Line (in red) of ``magic'' parameters $(\alpha,\beta)$, at which four exact flatbands (two per sublattice) appear. The two flatbands per sublattice have total Chern number $|C|=1$. At the point  where the red line crosses $\alpha$ axis ($\alpha=0.5279$ and $\beta=0$), the four flatbands (two per sublattice) corresponding to the red mBZ can be unfolded into two flatbands (one per sublattice) corresponding to the blue mBZ.
     (b, c) Many-body energy spectra at filling $\nu=2/3$ for $\alpha=0.5279,\beta=0$ and $\alpha=0.6651,\beta=0.7183$ respectively. Both show three quasi-degenerate ground states (marked by red circles). In both cases, the many-body Chern numbers of the ground states are $C_\text{mb}=1/3$. 
     (d) Many-body gap $\Delta E$ between the lowest energy excited state and highest energy ground state (among the three quasi-degnerate ground state) as the ratio $\beta/\alpha$ is varied along red line in (a).
    }
     \label{fig:s2}
\end{figure}
\newpage
\section{Appendix C. Particle entanglement spectra of the $\nu=2/3$ FCI states in {\color{black}Figs.~3(g) and~A2(b-c)}}
\begin{figure}[h!]
     \centering
\includegraphics[width=\linewidth]{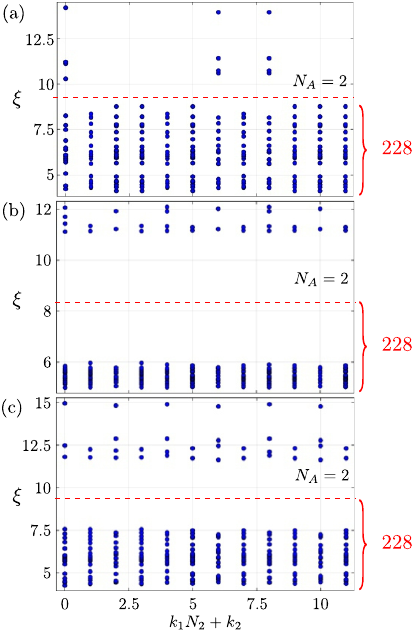}
     \caption{Particle entanglement spectra of the $\nu=2/3$ FCI ground-states in {\color{black}Figs.~3(g),~A2(b) and A2(c)} for a particle cut of $N_A=2$ and $N_B=6$ (total number of electrons $N_e=N_A+N_B=8$) are shown in (a), (b) and (c), respectively (see SM~\cite{SM2022} for details). In all three cases, there are gaps marked by red dashed lines, below which there are 228 states. This number matches with the PES counting of $\nu=1/3$ FCI state for $N_e=8$ electrons in a $C=1$ band with $N=24$ single particle states partitioned into two groups with $N_A=2$ and $N_B=6$ electrons. In (a) and (c), there are small gaps above the lowest 66 states too. We showed that in Fig.~\ref{fig:s2} that the ground states corresponding to (b) and (c) are adiabatically connected. The fact that low lying spectra of (a) and (c) have same counting indicates that FCI state in (a) is the same as that in (c).
    }
     \label{fig:em1}
\end{figure}
\newpage
\section{Appendix D. Particle entanglement spectra of the $\nu=4/3$ FCI states in {\color{black}Figs.~4(f,g)}}
\begin{figure}[h!]
     \centering
\includegraphics[width=\linewidth]{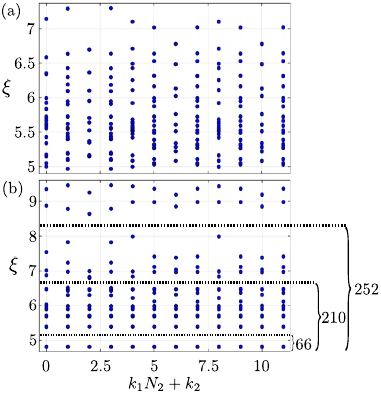}
     \caption{Particle entanglement spectra of the $\nu=4/3$ FCI ground-states in {\color{black}Figs.~4(f) and 4(g)} for a particle cut of $N_A=2$ and $N_B=6$ (total number of electrons $N_e=N_A+N_B=8$) are shown in (a) and (b), respectively. The PES in (a) does not show any gap. The counting of the low-lying states of PES in (b) matches with a ground state with one fully filled Wannier band and a $1/3$ filled Chern band (see {\color{black}SM}~\cite{SM2022} for details).
    }
     \label{fig:em1p}
\end{figure}

\clearpage
\let\addcontentsline\oldaddcontentsline
\onecolumngrid

\makeatletter
\renewcommand \thesection{S-\@arabic\c@section}
\renewcommand\thetable{S\@arabic\c@table}
\renewcommand \thefigure{S\@arabic\c@figure}
\renewcommand \theequation{S\@arabic\c@equation}
\makeatother
\setcounter{equation}{0}  
\setcounter{figure}{0}  
\setcounter{section}{0}  
\counterwithin{figure}{section} 
{
    \center \bf \large 
    Supplemental Material\vspace*{0.1cm}\\ 
    \vspace*{0.0cm}
}
\maketitle
\tableofcontents

\section{Symmetries of the single-particle Hamiltonian}
The Hamiltonian in the main text has the form
\begin{equation}\label{eq:hamiltonian1}
     \mathcal{H}(\mathbf{r})
      =\begin{pmatrix}\mathbf{0} &\mathcal{D}^\dagger(\mathbf{r})\\\mathcal{D}(\mathbf{r}) & \mathbf{0}\end{pmatrix},  \mathcal{D}(\mathbf{r}) = \mathcal{D}_K(\mathbf{r})+\mathcal{D}_U(\mathbf{r}) = -4 \overline{\partial_{z}}^2 +\tilde{A}(\mathbf{r}),
 \end{equation}
near $\Gamma$ point in the Brillouin zone (BZ) of the constituent material having a QBCP, at $\Gamma$, protected by $n$-fold rotation symmetry $\mathcal{C}_{nz}$ ($n=3$ or $4$) and time reversal symmetry $\mathcal{T}$. The operator $\mathcal{D}(\mathbf{r})$ has kinetic part $\mathcal{D}_K(\mathbf{r})$ consisting of spatial derivatives and potential term $D_U(\mathbf{r};\boldsymbol{\alpha})$ which arises from moir\'e periodic potential (in our case, the potential is a strain field). The kinetic part is antiholomorphic: $\mathcal{D}_K(\mathbf{r}) = (-2i\overline{\partial_{z}})^2 $, where $z = x+ i y$, the overline stands for complex conjugation. We require that the rotation symmetry $\mathcal{C}_{nz}$ that protects the QBCP is not broken by the moir\'e potential. This implies that under $\mathcal{C}_{nz}$ rotation, $\mathcal{D}(\mathbf{r}) = \mathcal{D}_k(\mathbf{r})+\mathcal{D}_U(\mathbf{r};\boldsymbol{\alpha})$ transforms as $\mathcal{D}(\mathcal{C}_{nz}\mathbf{r}) =(\omega^*)^2 \mathcal{D}(\mathbf{r})$, where $\omega = e^{10\pi i/n}$. Consequently, the Hamiltonian satisfies 
$\text{Diag}\{\omega,\overline{\omega}\} \mathcal{H}(\mathbf{r})\text{Diag}\{\overline{\omega},\omega\} = \mathcal{H}(\mathcal{C}_{nz}\mathbf{r})$. Furthermore, this Hamiltonian has time reversal symmetry $\mathcal{T}$ such that $\sigma_x \mathcal{H}^*(\mathbf{r})\sigma_x = \mathcal{H}(\mathbf{r})$. Due to the off-diagonal nature of the Hamiltonian, it also has chiral symmetry $\mathcal{S}$ such that $\sigma_z \mathcal{H}(\mathbf{r})\sigma_z = -\mathcal{H}(\mathbf{r})$. Moreover, in case of a $\mathcal{C}_{6z}$ symmetric Hamiltonian, the $\mathcal{C}_{2z}$ constrains the Hamiltonian to satisfy $\mathcal{H}(-\mathbf{r}) = \mathcal{H}(\mathbf{r})$, where the representation of $\mathcal{C}_{2z}$ is $\mathds{1}$. Note that we could have equally well chosen the $\mathcal{C}_{2z}$ representation to be $-\mathds{1}$. There can further be mirror symmetries $\mathcal{M}_{x/y}$ depending on the choice of $\tilde{A}(\mathbf{r})$, which are both represented as $\sigma_x$, and they constrain the Hamiltonian to satisfy $\sigma_x \mathcal{H}(\mathbf{r})\sigma_x = \mathcal{H}(\mathcal{M}_{x/y}\mathbf{r})$.

\section{Exact moir\'e flat band wave functions and their analogy to Landau level wavefunctions}

In this section, we construct the flat band wavefunctions for the four flat bands shown in {\color{red}Figs.1 and 3} of the main text. However, before describing that we review the construction of wave-functions for the case two flat bands (one per sublattice).

\subsection{Construction of wave-functions in the case of two flat bands (single flat band per sublattice) and similarity to the lowest Landau level wavefunction}
An exact FB of $\mathcal{H}(\mathbf{r})$ (Eq.~\eqref{eq:hamiltonian1}) at energy $E= 0$ with wave-function $\Psi_\mathbf{k}(\mathbf{r})$ satisfies $\mathcal{H}(\mathbf{r})\Psi_\mathbf{k}(\mathbf{r}) = \mathbf{0}$ for all $\mathbf{k}$. The construction of such a $\Psi_\mathbf{k}(\mathbf{r})$ is as follows. Note that due to $\mathcal{C}_{nz}$ and chiral symmetry $\mathcal{S} = \sigma_z\otimes \mathds{1}$, the two fold degeneracy of QBCP at $\Gamma$ remains at $E= 0$ for any $\mathcal{D}_U (\mathbf{r})$ that keeps $\mathcal{C}_{nz}$ symmetry. This means that there are always two sublattice-polarized wave-functions $\Psi_{\Gamma,1}(\mathbf{r}) = \{\psi_{\Gamma}(\mathbf{r}),0\}$ and $\Psi_{\Gamma,2}(\mathbf{r}) = \{0,\psi_{\Gamma}^*(\mathbf{r})\}$ satisfying $\mathcal{H}(\mathbf{r})\Psi_{\Gamma,i}(\mathbf{r})=\mathbf{0}$, or equivalently $\mathcal{D}(\mathbf{r})\psi_{\Gamma}(\mathbf{r})=\mathbf{0}$. If there are exact flatbands, the wave-functions can be written as $\{\psi_\mathbf{k}(\mathbf{r}),0\}$ and $\{0,\overline{\psi_{-\mathbf{k}}(\mathbf{r})}\}$. Since the kinetic part $\mathcal{D}_K(\mathbf{r})$ is antiholomorphic, the trial wavefunction is naturally $\psi_{\mathbf{k}}(\mathbf{r}) = f_\mathbf{k}(z)\psi_{\Gamma}(\mathbf{r})$, where $f_\mathbf{k}(z)$ is a holomorphic function satisfying $\overline{\partial_{z}}f_\mathbf{k}(z) = 0$. The function $f_\mathbf{k}(z)$ needs to satisfy Bloch periodicity (translation by moir\'e lattice vector $\mathbf{a}$ gives phase shift $e^{i\mathbf{k}\cdot\mathbf{a}}$). However, from Louiville's theorem, such a holomorphic function must have poles, making $\psi_\mathbf{k}(\mathbf{r})$ divergent, unless $\psi_{\Gamma}(\mathbf{r})$ has a zero that cancels the pole. Conversely, if $\psi_{\Gamma}(\mathbf{r})$ has zero at $\mathbf{r}_0$, a Bloch periodic holomorphic function
\begin{equation}
\label{eq:fkz}
\begin{split}
    &f_\mathbf{k}(z;\mathbf{r}_0)  = e^{i (\mathbf{k}\cdot\mathbf{a}_1) z/a_1}\frac{\vartheta\left(\frac{z-z_0}{a_1}-\frac{k}{b_2},\tau\right)}{\vartheta\left(\frac{z-z_0}{a_1},\tau\right)}= e^{i\mathbf{k}\cdot\mathbf{r}} \tilde{f}_\mathbf{k}(\mathbf{r};\mathbf{r}_0),\\
    & \tilde{f}_\mathbf{k}(\mathbf{r};\mathbf{r}_0) = e^{-i(\mathbf{b}_2\cdot\mathbf{r})k/b_2}\frac{\vartheta\left(\frac{z-z_0}{a_1}-\frac{k}{b_2},\tau\right)}{\vartheta\left(\frac{z-z_0}{a_1},\tau\right)}
\end{split}
\end{equation}
with a pole at $\mathbf{r}_0$ can be constructed. Here $\vartheta (z,\tau) = -i\sum_{n=-\infty}^\infty(-1)^n e^{\pi i \tau(n+1/2)^2+\pi i(2n+1)z}$ is the Jacobi theta function of the first type~\cite{ledwith2020fractional}, $\mathbf{a}_i$ are lattice vectors, $\mathbf{b}_i$ are the corresponding reciprocal lattice vectors ($\mathbf{a}_i\cdot\mathbf{b}_j = 2\pi \delta_{ij}$), $a_i = (\mathbf{a}_i)_x+ i (\mathbf{a}_i)_y$, $b_i = (\mathbf{b}_i)_x+ i (\mathbf{b}_i)_y$, $z_0 = (\mathbf{r}_0)_x+ i (\mathbf{r}_0)_y$, $k = k_x + i k_y$, and $\tau = a_2/a_1$. Using the relations $\theta(z+1,\tau) = -\theta(z,\tau)$ and $\theta(z+\tau,\tau) = -\exp(-2\pi i z -\pi i \tau)\theta(z,\tau)$, it can be checked that $f_\mathbf{k}(z;\mathbf{r}_0)$ is indeed Bloch periodic. Furthermore, since $\theta(z,\tau)$ has a pole at $z=m+n\tau$ (for all integer $m,n$), $f_\mathbf{k}(z;\mathbf{r}_0)$ has poles at $\mathbf{r} = \mathbf{r}_0+m \mathbf{a}_1+n\mathbf{a}_2$ and zeros at $\mathbf{r} = \mathbf{r}_0-\frac{1}{2\pi}(\hat{z}\cdot\mathbf{a}_1\times\mathbf{a}_2)(\hat{z}\times\mathbf{k})+m \mathbf{a}_1+n\mathbf{a}_2$. Remarkably, at ``magic'' values of strain strength $\alpha$, the wave-function $\psi_{\Gamma}(\mathbf{r})$ has a zero, which allows for such $f_\mathbf{k}(z;\mathbf{r}_0)$, and in turn gives rise to two exact FBs. For completeness, we write the expression of $\psi_\mathbf{k}(\mathbf{r})$ below
\begin{equation}\label{eq:singleWF}
     \psi_\mathbf{k}(\mathbf{r}) = e^{i (\mathbf{k}\cdot\mathbf{a}_1) z/a_1}\vartheta\left(\frac{z-z_0}{a_1}-\frac{k}{b_2},\tau\right)\frac{\psi_\Gamma(\mathbf{r})}{\vartheta\left(\frac{z-z_0}{a_1},\tau\right)}.
\end{equation}
Furthermore, the periodic part $\tilde{f}_\mathbf{k}(\mathbf{r};\mathbf{r}_0)$ is a holomorphic function of $k$: $\overline{\partial_{k}}\tilde{f}_\mathbf{k}(\mathbf{r};\mathbf{r}_0) = \frac{1}{2}[(\partial_{k_x}+i\partial_{k_y})]\tilde{f}_\mathbf{k}(\mathbf{r};\mathbf{r}_0) = 0$ as can be seen from Eq.~\eqref{eq:fkz}~\cite{kharchev2015theta,ledwith2020fractional}; this property along with the presence of the zero in the wave-function can be used to prove that  wave functions of this form carry Chern number $C = \pm 1$ (see~\cite{wang2021exact} for a proof). 

An important point to note here is that not all zeros of of the wavefunction $\psi_\Gamma(\mathbf{r})$ can be used to construct $\psi_\mathbf{k}(\mathbf{r})$~\cite{sarkar2023symmetry}. The reason is that generically near a zero at $\mathbf{r}_0$ of $\psi_\Gamma(\mathbf{r})$, the wavefunction has the form $\psi_\Gamma(\mathbf{r}_0+\epsilon(x,y))\sim \epsilon(ax+iby)+\mathcal{O}(\epsilon^2)$ where $a,b$ are some complex numbers. Hence, the total wavefunction $\psi_\mathbf{k}(\mathbf{r})$ near the zero behaves as $\psi_\mathbf{k}(\mathbf{r}_0+\epsilon(x,y))\sim \frac{(ax+iby)}{c(x+iy)}+\mathcal{O}(\epsilon)$, where $c$ is a complex number. This function is generically singular at $x = y= 0$, and hence not a good wavefunction. It was shown in~\cite{sarkar2023symmetry} that unless fine-tuned, only zeros of $\psi_\Gamma(\mathbf{r})$ at high symmetry points with $\mathcal{C}_{nz}$ (where $n\geq 2$) symmetry of the moir\'e unit cell can be used to construct flatband wavefunction $\psi_\mathbf{k}(\mathbf{r})$. The reason is that near a zero at a high symmetry point with $\mathcal{C}_{nz}$ (where $n\geq 2$) symmetry, $\psi_\Gamma(\mathbf{r})$ behaves as $\psi_\Gamma(\mathbf{r}_0+\epsilon(x,y))= \epsilon^2(c_1(x+iy)^2+c_2(x^2+y^2))+\mathcal{O}(\epsilon^3) = \epsilon^2 (c_1z^2+c_2z\overline{z})+\mathcal{O}(\epsilon^3)$ (where $c_1,c_2$ are complex numbers) for single layer QBCP models with moir\'e potential, and hence $\psi_\mathbf{k}(\mathbf{r}_0+\epsilon(x,y))\sim \epsilon (c_1z+c_2\overline{z})/c +\mathcal{O}(\epsilon^2)$ is a smooth function.

Note that these wave-functions are just like lowest Landau Level (LLL) wave-functions on a torus, which are also constructed using theta functions~\cite{haldane1985periodic}. To see this, let us consider the Hamiltonian for 2D free electrons under out-of-plane magnetic field
\begin{equation}\label{eq:LLHam}
     \mathcal{H}_\text{LL}(\mathbf{r}) = \frac{\hbar^2}{2m}\left[\left(-i\partial_x-\frac{e}{\hbar}A_x(\mathbf{r})\right)^2+\left(-i\partial_y-\frac{e}{\hbar}A_y(\mathbf{r})\right)^2\right]
\end{equation}
where $\mathbf{A}(\mathbf{r})$ is vector potential, $m$ is the electron mass, $e$ is the charge of the electron. For a constant magnetic field $\mathbf{B} =B_0\hat{z}$, we choose the following Landau gauge $\mathbf{A}(\mathbf{r}) = B_0(\hat{b}_2\cdot(\mathbf{r}-\mathbf{r}_0))(\hat{z}\times\hat{b}_2)$, where $\hat{b}_2 = \mathbf{b}_2/|\mathbf{b}_2|$ is the unit vector along the reciprocal lattice vector $\mathbf{b}_2$.
Plugging the ansatz for Landau level eigenstate of this Hamiltonian to be $\psi^\text{LLL}(\mathbf{r})=g(\mathbf{r})\exp\left(-\frac{1}{2\ell_B^2}(\hat{b}_2\cdot(\mathbf{r}-\mathbf{r}_0))^2\right)$ (here $\ell_B^2 = \hbar/(eB_0)$), we find
\begin{equation}
     \frac{\hbar^2}{2m}\left[-4\partial_z+\frac{4}{\ell_B^2}(\hat{b}_2\cdot(\mathbf{r}-\mathbf{r}_0))(\hat{b}_{2x}-i\hat{b}_{2y})\right]\overline{\partial_z} g(\mathbf{r}) = (E-\frac{1}{2}\hbar\omega_c)g(\mathbf{r}),
\end{equation}
where $\omega_c = eB_0/m$ is cyclotron frequency. Then, if $\overline{\partial_z} g(\mathbf{r}) = 0$, $E= \frac{1}{2}\hbar\omega_c$ is the lowest Landau level energy. This implies for LLL wavefunctions, $g$ is a holomorphic function of $z$. Next, on a tours magnetic flux through a magnetic unit cell (given by $\mathbf{a}_1$ and $\mathbf{a}_2$) is $2\pi$ in the units of $\hbar/e$ (one flux quantum per unit cell), or in other words $2\pi \ell_B^2 = \hat{z}\cdot(\mathbf{a}_1\times\mathbf{a}_2)$. The wavefunctions $\psi^\text{LLL}_\mathbf{k}(\mathbf{r})$ satisfy magnetic Bloch-periodicity
\begin{equation}
     \psi^\text{LLL}_\mathbf{k}(\mathbf{r}+\mathbf{a}_1) = -e^{i\mathbf{k}\cdot\mathbf{a}_1}\psi^\text{LLL}_\mathbf{k}(\mathbf{r}),\,\psi^\text{LLL}_\mathbf{k}(\mathbf{r}+\mathbf{a}_2) = -e^{i\mathbf{k}\cdot\mathbf{a}_2}e^{-2\pi i\frac{\mathbf{a}_1\cdot(\mathbf{r}-\mathbf{r}_0+\mathbf{a}_2/2)}{|\mathbf{a}_1|^2}}\psi^\text{LLL}_\mathbf{k}(\mathbf{r}),
\end{equation}
where the constant part in the second exponential in the second equation is a gauge choice. Using the definition and properties of Jacobi theta function mentioned above, it can be verified that
\begin{equation}\label{eq:LLLWF}
     \psi^\text{LLL}_\mathbf{k}(\mathbf{r}) = e^{i(\mathbf{k}\cdot\mathbf{a}_1)z/a_1}\vartheta\left(\frac{z-z_0}{a_1}-\frac{k}{b_2},\tau\right) \exp\left(-\frac{\pi}{\hat{z}\cdot(\mathbf{a}_1\times\mathbf{a}_2)}(\hat{b}_2\cdot(\mathbf{r}-\mathbf{r}_0))^2\right).
\end{equation}
This expression is very similar to the one in Eq.~\eqref{eq:singleWF}; the only difference is in the $\mathbf{k}$ independent factors in the functions $\frac{\psi_\Gamma(\mathbf{r})}{\vartheta\left(\frac{z-z_0}{a_1},\tau\right)}$ vs $\exp\left(-\frac{\pi}{\hat{z}\cdot(\mathbf{a}_1\times\mathbf{a}_2)}(\hat{b}_2\cdot(\mathbf{r}-\mathbf{r}_0))^2\right)$. Using this analogy between LLL wavefunction and the moir\'e flatband wavefunction, we can rewrite the moir\'e flatband wavefunction as
\begin{equation}
\begin{split}
     &\psi_\mathbf{k}(\mathbf{r})\\
     &= \psi^\text{LLL}_\mathbf{k}(\mathbf{r})\frac{\psi_\Gamma(\mathbf{r})}{\psi^\text{LLL}_{\Gamma}(\mathbf{r})}\\
     &=e^{i(\mathbf{k}\cdot\mathbf{a}_1)z/a_1}\vartheta\left(\frac{z-z_0}{a_1}-\frac{k}{b_2},\tau\right)e^{\left(-\frac{\pi}{\hat{z}\cdot(\mathbf{a}_1\times\mathbf{a}_2)}(\hat{b}_2\cdot(\mathbf{r}-\mathbf{r}_0))^2\right)} \left|\frac{\psi_\Gamma(\mathbf{r})}{\psi^\text{LLL}_{\Gamma}(\mathbf{r})}\right|\zeta(\mathbf{r})\\
     &=h_\mathbf{k}(z) e^{-K(\mathbf{r})/4\ell_B^2}\zeta(\mathbf{r}) =\phi_\mathbf{k}(\mathbf{r})\zeta(\mathbf{r}),
\end{split}
\end{equation}
where $\phi_\mathbf{k}(\mathbf{r})e^{K(\mathbf{r})/4\ell_B^2} =h_\mathbf{k}(z) = e^{i(\mathbf{k}\cdot\mathbf{a}_1)z/a_1}\vartheta\left(\frac{z-z_0}{a_1}-\frac{k}{b_2},\tau\right)$ is a holomorphic function of $z$, $e^{-K(\mathbf{r})/4\ell_B^2} = |\psi_\Gamma(\mathbf{r})/\vartheta\left(\frac{z-z_0}{a_1},\tau\right)|$, and $\zeta(\mathbf{r}) = e^{i\arg\left(\psi_\Gamma(\mathbf{r})/\vartheta\left(\frac{z-z_0}{a_1},\tau\right)\right)}$. The first part of this expression, $h_\mathbf{k}(z)e^{-K(\mathbf{r})/4\ell_B^2}$ has the interpretation of LLL wavefunction in spatially varying magnetic field $B(\mathbf{r}) = B_0\partial_z\overline{\partial_z}K(\mathbf{r})$~\cite{ledwith2020fractional}. Indeed, the LLL wavefunction we wrote in Eq.~\eqref{eq:LLLWF} has the form of  $\psi^\text{LLL}_\mathbf{k}(\mathbf{r}) = h_\mathbf{k}(z) e^{-K_0(\mathbf{r})/4\ell_B^2}$, with $K_0(\mathbf{r}) = 2(\hat{b}_2\cdot(\mathbf{r}-\mathbf{r}_0))^2$ and magnetic field $B(\mathbf{r})= B_0\partial_z\overline{\partial_z}K(\mathbf{r}) = B_0/4 (\partial_x^2+\partial_y^2) 2(\hat{b}_2\cdot(\mathbf{r}-\mathbf{r}_0))^2 = B_0$. In the case of moir\'e flatband wavefunction, the function $K(\mathbf{r})$ can be split into two parts
\begin{equation}
     K(\mathbf{r}) = K_0(\mathbf{r})+K_\lambda(\mathbf{r}), 
\end{equation}
where $K_0(\mathbf{r})$ acounts for the average magnetic field $B_0$, and $K_\lambda(\mathbf{r})$ acounts for the spatial variation of the magnetic field about the mean. Explicitly written, 
\begin{equation}
     K_\lambda(\mathbf{r}) = 2\ell_B^2 \log \left|\frac{\psi^\text{LLL}_\Gamma(\mathbf{r})}{\psi_\Gamma(\mathbf{r})}\right|^2 = \frac{\hat{z}\cdot(\mathbf{a}_1\times\mathbf{a}_2)}{\pi}\log \left|\frac{\psi^\text{LLL}_\Gamma(\mathbf{r})}{\psi_\Gamma(\mathbf{r})}\right|^2, \text{ and } \frac{B(\mathbf{r})}{B_0} = 1+  \frac{\hat{z}\cdot(\mathbf{a}_1\times\mathbf{a}_2)}{\pi}\partial_z\overline{\partial_z}\log \left|\frac{\psi^\text{LLL}_\Gamma(\mathbf{r})}{\psi_\Gamma(\mathbf{r})}\right|^2.
\end{equation}
It is worth mentioning that since $\psi^\text{LLL}_\Gamma(\mathbf{r})$ satisfies magnetic translation symmetry and $\psi_\Gamma(\mathbf{r})$ satisfies ordinary translation symmetry, $K_\lambda(\mathbf{r})$ is a periodic function with moir\'e periodicity. Hence, the magnetic flux of $K_\lambda(\mathbf{r})$ is zero in a moir\'e unit cell, and total magnetic flux of $K(\mathbf{r})$ is $2\pi$ in the units of $\hbar/e$.

The ``fracionalization'' of $\psi_\mathbf{k}(\mathbf{r})$ into LLL wavefunction in a varying magnetic field $\phi_\mathbf{k}(\mathbf{r})$ and $\zeta(\mathbf{r})$ was used in~\cite{dong2023composite} as a parton-ansatz for the electrons in moir\'e flatbands. In the case twisted bilayer graphene or twisted bilayer MoTe$_2$, $\zeta(\mathbf{r})$ is a vector with components indexed by the layer numbers, whereas in our case, where there is a single layer, $\zeta(\mathbf{r})$ is just a phase as was written earlier. Furthermore, since $\phi_\mathbf{k}(\mathbf{r})$ is a LLL wavefunction, it satisfies magnetic-translation symmetry, and hence $\zeta(\mathbf{r})$ satisfies magnetic translation symmetry with a magnetic field of opposite sign (since the total wavefunction $\psi_\mathbf{k}(\mathbf{r})$ satisfies ordniary translation symmmetry). The magnetic translation symmetry implies a vortex like winding of the phase of $\zeta(\mathbf{r})$ by $-2\pi$ around the moir\'e unit cell~\cite{guerci2024layer}. For multilayer system this winding can be accomodated by a layer skyrmion structure of $\zeta(\mathbf{r})$; however, in a single layer system this winding implies a singularity in $\zeta(\mathbf{r})$, or in other words a zero in $\frac{\psi_\Gamma(\mathbf{r})}{\vartheta\left(\frac{z-z_0}{a_1},\tau\right)}$ of type $\overline{z}$. Indeed, some of the authors of this article showed in~\cite{sarkar2023symmetry}, that for a single layer QBCP system with moir\'e potential, the Taylor expansion of $\psi_\Gamma(\mathbf{r})$ near its zero $\mathbf{r}_0$ at a high symmetry point in the unit cell is of the form $\psi_\Gamma(\mathbf{r}_0+\epsilon(x,y))\sim \epsilon^2(x^2+y^2)+\mathcal{O}(\epsilon^3) = \epsilon^2z\overline{z}+\mathcal{O}(\epsilon^3)$, and hence $\frac{\psi_\Gamma(\mathbf{r}_0+\epsilon(x,y))}{\vartheta\left(\frac{x+iy}{a_1},\tau\right)}\sim(x-iy)=\bar{z}$ near the zero (since $\vartheta\left(\frac{x+iy}{a_1},\tau\right)\sim (x+iy)$ near $x=y=0$). 

\underline{\textit{Singularities in the magnetic field distribution}:} Below we show that the magnetic field distribution $B(\mathbf{r})$ for an exact flat band in single layer system with QBCP + moir\'e potential has singularities at the point where $\psi_\Gamma(\mathbf{r})$ has a zero. 

To this end, first note that it was shown in~\cite{sarkar2023symmetry} that the wavefunction $\psi_\Gamma(\mathbf{r})$ near its zero at a high symmetry point in the unit cell $\mathbf{r}_0$ has the form $\psi_\Gamma(\mathbf{r}_0+\epsilon(x,y)) = \epsilon^2(c_1(x+iy)^2+c_2(x^2+y^2)) +\mathcal{O}(\epsilon^3) = \epsilon^2(c_1z^2+c_2z\overline{z}) +\mathcal{O}(\epsilon^3)$ (where $c_2$ is some complex number) due to rotation symmetry around $\mathbf{r}_0$:  $\psi_\Gamma(\mathbf{r}_0+\mathcal{C}_{nz}\mathbf{r}) = \psi_\Gamma(\mathbf{r}_0+\mathbf{r})$, where $\mathcal{C}_{nz}$ is the $n$-fold rotation symmetry with $n=2,3,4,6$. Furthermore, for $\mathcal{C}_{3z}$, $\mathcal{C}_{4z}$ and $\mathcal{C}_{6z}$ symmetric points, $c_1=0$; $c_1$ can only be nonzero near a $\mathcal{C}_{2z}$ symmetric point.

Next we expand $\psi_\Gamma(\mathbf{r}_0+\epsilon(x,y))$ to order $\epsilon^3$. There are four possible terms: $z^3,z^2\overline{z},z\overline{z}^2,\overline{z}^3$. If $\mathbf{r}_0$ is a $\mathcal{C}_{3z}$ center, only $z^3$ and $\overline{z}^3$ are allowed, since only these two terms are $\mathcal{C}_3$ invariant. However, it was shown in~\cite{sarkar2023symmetry} that the coefficient of $\overline{z}^3$ becomes zero at the magic value of the moir\'e potential amplitude for which the exact flat band appears. It is also clear that if the term $\overline{z}^3$ were there, then $\frac{\psi_\Gamma(\mathbf{r}_0+\epsilon(x,y))}{\vartheta\left(\frac{x+iy}{a_1},\tau\right)} \sim \dots+ \overline{z}^3/z+\dots$ would be singular (and not a good wave function) since $\overline{z}^3/z$ is singular. Hence, if $\mathbf{r}_0$ is $\mathcal{C}_{3z}$ symmetric, the wavefunction $\psi_\Gamma$ near $\mathbf{r}_0$ behaves as $\psi_\Gamma(\mathbf{r}_0+\epsilon(x,y)) = \epsilon^2c_2z\overline{z} + \epsilon^3 c_3 z^3+\mathcal{O}(\epsilon^4)$ (where $c_2,c_3$ are complex numbers). On the other hand, if $\mathbf{r}_0$ has $\mathcal{C}_{2z}$ symmetry, the cubic terms are not allowed since none of $z^3,z^2\overline{z},z\overline{z}^2,\overline{z}^3$ is $\mathcal{C}_{2z}$ symmetric. Hence, $\psi_\Gamma(\mathbf{r}_0+\epsilon(x,y)) = \epsilon^2c_2z\overline{z} +\mathcal{O}(\epsilon^4)$ if $\mathbf{r}_0$ is a $\mathcal{C}_{4z}$ or $\mathcal{C}_{6z}$ center, whereas $\psi_\Gamma(\mathbf{r}_0+\epsilon(x,y)) = \epsilon^2(c_1z^2+c_2z\overline{z}) +\mathcal{O}(\epsilon^4)$ if $\mathbf{r}_0$ is a $\mathcal{C}_{2z}$ center. 

Next, consider the behavior of $\psi_\Gamma^\text{LLL}(\mathbf{r})$ near $\mathbf{r}_0$. We find $\vartheta\left(\frac{\epsilon(x+iy)}{a_1},\tau\right)= \epsilon c_1'z+\mathcal{O}(\epsilon^3)$ (where $c_1'$ is a complex number). Furthermore, the factor $e^{\left(-\frac{\pi}{\hat{z}\cdot(\mathbf{a}_1\times\mathbf{a}_2)}(\hat{b}_2\cdot(\mathbf{r}-\mathbf{r}_0))^2\right)} = e^{\left(-\frac{\pi}{\hat{z}\cdot(\mathbf{a}_1\times\mathbf{a}_2)}(\hat{b}_2\cdot(\epsilon(x,y)))^2\right)} = 1+\mathcal{O}(\epsilon^2)$. Hence, $\psi_\Gamma^\text{LLL}(\mathbf{r}_0+\epsilon(x,y)) = \epsilon c_1' z +\mathcal{O}(\epsilon^3)$. Putting all of it together, we get
\begin{eqnarray}
     \frac{\psi_\Gamma(\mathbf{r}_0+\epsilon(x,y))}{\psi^\text{LLL}_\Gamma(\mathbf{r}_0+\epsilon(x,y))} = \begin{cases}
          \epsilon(\frac{c_1}{c_1'}z+\frac{c_2}{c_1'}\overline{z}) + \mathcal{O}(\epsilon^3), &\text{ if $\mathbf{r}_0$ is $\mathcal{C}_{2z}$ symmetric}\\
          \epsilon\frac{c_2}{c_1'}\overline{z} + \epsilon^2 \frac{c_3}{c_1'}z^2 +\mathcal{O}(\epsilon^3), &\text{ if $\mathbf{r}_0$ is $\mathcal{C}_{3z}$ symmetric}\\
          \epsilon\frac{c_2}{c_1'}\overline{z} + \mathcal{O}(\epsilon^3), &\text{ if $\mathbf{r}_0$ is $\mathcal{C}_{4z}$ or $\mathcal{C}_{6z}$ symmetric},
     \end{cases}
\end{eqnarray}
near the zero $\mathbf{r}_0$. Hence, near the zero, $K_\lambda(\mathbf{r})$ behaves as
\begin{equation}
\begin{split}
     K_\lambda(\mathbf{r}_0+\epsilon(x,y)) &= - 2\ell_B^2\log \left| \frac{\psi_\Gamma(\mathbf{r}_0+\epsilon(x,y))}{\psi^\text{LLL}_\Gamma(\mathbf{r}_0+\epsilon(x,y))}\right|^2\\ 
     &= \begin{cases}
          -2\ell_B^2 \log(\epsilon^2|c_1/c_1'|^2) -2\ell_B^2 \log\left|z+\frac{c_2}{c_1}\overline{z}\right|^2-2\ell_B^2 \log\left|1+\mathcal{O}(\epsilon^2)\right|^2, &\text{ if $\mathbf{r}_0$ is $\mathcal{C}_{2z}$ symmetric}\\
          -2\ell_B^2 \log(\epsilon^2|c_2/c_1'|^2)-2\ell_B^2\log(z\overline{z})-2\ell_B^2 \log\left|1+\epsilon\frac{c_3z^2}{c_2\overline{z}}+\mathcal{O}(\epsilon^2)\right|^2, &\text{ if $\mathbf{r}_0$ is $\mathcal{C}_{3z}$ symmetric}\\
          -2\ell_B^2 \log(\epsilon^2|c_2/c_1'|^2)-2\ell_B^2\log(z\overline{z})-2\ell_B^2 \log\left|1+\mathcal{O}(\epsilon^2)\right|^2, &\text{ if $\mathbf{r}_0$ is $\mathcal{C}_{4z}$/$\mathcal{C}_{6z}$ symmetric}\\
     \end{cases}\\
     &=\begin{cases}
          -2\ell_B^2 \log(\epsilon^2|c_1/c_1'|^2) -2\ell_B^2 \log\left|z+\tilde{c}\overline{z}\right|^2+\mathcal{O}(\epsilon^2), &\text{ if $\mathbf{r}_0$ is $\mathcal{C}_{2z}$ symmetric}\\
          -2\ell_B^2 \log(\epsilon^2|c_2/c_1'|^2)-2\ell_B^2\log(z\overline{z})-2\ell_B^2\epsilon \left(\tilde{c}_1\frac{z^2}{\overline{z}}+\overline{\tilde{c}_1}\frac{\overline{z}^2}{z}\right)+\mathcal{O}(\epsilon^2), &\text{ if $\mathbf{r}_0$ is $\mathcal{C}_{3z}$ symmetric}\\
          -2\ell_B^2 \log(\epsilon^2|c_2/c_1'|^2)-2\ell_B^2\log(z\overline{z})+\mathcal{O}(\epsilon^2), &\text{ if $\mathbf{r}_0$ is $\mathcal{C}_{4z}$/$\mathcal{C}_{6z}$ symmetric}\\
     \end{cases},%
\end{split}
\end{equation}
where $\tilde{c} = c_2/c_1$ and $\tilde{c}_1 = c_3/c_2$. This implies that near the zero the magnetic field behaves as 
\begin{align}\label{eq:Bsingularity}
     B(\mathbf{r}) -B_0 &= B(\mathbf{r}_0+\epsilon(x,y)) -B_0\nonumber\\  &=  \begin{cases}
          -2B_0\ell_B^2 \partial_z\overline{\partial_z} \log\left|z+\tilde{c}\overline{z}\right|^2+\mathcal{O}(\epsilon^2), &\text{ if $\mathbf{r}_0$ is $\mathcal{C}_{2z}$ symmetric}\\
          -2B_0\ell_B^2 \partial_z\overline{\partial_z} \log(z\overline{z})-2B_0\ell_B^2\epsilon \partial_z\overline{\partial_z}\left(\tilde{c}_1\frac{z^2}{\overline{z}}+\overline{\tilde{c}_1}\frac{\overline{z}^2}{z}\right)+\mathcal{O}(\epsilon^2), &\text{ if $\mathbf{r}_0$ is $\mathcal{C}_{3z}$ symmetric}\\
          -2B_0\ell_B^2 \partial_z\overline{\partial_z} \log(z\overline{z})+\mathcal{O}(\epsilon^2), &\text{ if $\mathbf{r}_0$ is $\mathcal{C}_{4z}$/$\mathcal{C}_{6z}$ symmetric}\\
     \end{cases}\nonumber\\
     \nonumber\\
     \nonumber\\
     &= - \frac{B_0\ell_B^2}{2}(\partial_x^2+\partial_y^2)\log(x^2+y^2)\nonumber\\
     &\phantom{=}\;-2B_0\ell_B^2 \begin{cases}
          \frac{1}{4|z|^2}\partial_\phi^2 \log(1+|\tilde{c}|^2+2|\tilde{c}|\cos(2\phi-\phi_0))+\mathcal{O}(\epsilon^2), &\text{ if $\mathbf{r}_0$ is $\mathcal{C}_{2z}$ symmetric}\\
          -4\epsilon\left(\tilde{c}_1\frac{z}{\overline{z}^2}+\overline{\tilde{c}_1}\frac{\overline{z}}{z^2}\right)+\mathcal{O}(\epsilon^2), &\text{ if $\mathbf{r}_0$ is $\mathcal{C}_{3z}$ symmetric}\\
          \mathcal{O}(\epsilon^2), &\text{ if $\mathbf{r}_0$ is $\mathcal{C}_{4z}$/$\mathcal{C}_{6z}$ symmetric}\\
     \end{cases}\nonumber\\
     &= -\frac{B_0\ell_B^2}{2} 4\pi\epsilon((x,y))+\text{finite terms}\nonumber\\
     &\phantom{= -\frac{B_0\ell_B^2}{2} 4\pi\epsilon((x,y))}-2B_0\ell_B^2 \begin{cases}
          -\frac{2|\tilde{c}|(2|\tilde{c}|+(1+|\tilde{c}|^2)\cos(2\phi-\phi_0))}{|z|^2(1+|\tilde{c}|^2+2|\tilde{c}|\cos(2\phi-\phi_0))^2}&\text{ if $\mathbf{r}_0$ is $\mathcal{C}_{2z}$ symmetric}\\
          -4\epsilon\left(|\tilde{c}_1|\frac{e^{i(3\phi+\phi_1)}}{|z|}+|\tilde{c}_1|\frac{e^{-i(3\phi+\phi_1)}}{|z|}\right), &\text{ if $\mathbf{r}_0$ is $\mathcal{C}_{3z}$ symmetric}\\
          0, &\text{ if $\mathbf{r}_0$ is $\mathcal{C}_{4z}$/$\mathcal{C}_{6z}$ symmetric}\\
     \end{cases}\nonumber\\
     &= -\frac{2\pi\hbar}{e}\delta(\mathbf{r}-\mathbf{r}_0)+\text{finite terms}+\begin{cases}
          4\frac{\hbar}{e}\frac{|\tilde{c}|(2|\tilde{c}|+(1+|\tilde{c}|^2)\cos(2\phi-\phi_0))}{|\mathbf{r}-\mathbf{r}_0|^2(1+|\tilde{c}|^2+2|\tilde{c}|\cos(2\phi-\phi_0))^2}&\text{ if $\mathbf{r}_0$ is $\mathcal{C}_{2z}$ symmetric}\\
          16\frac{\hbar}{e}\epsilon\frac{|\tilde{c}_1|}{|\mathbf{r}-\mathbf{r}_0|}\cos(3\phi+\phi_1), &\text{ if $\mathbf{r}_0$ is $\mathcal{C}_{3z}$ symmetric}\\
          0, &\text{ if $\mathbf{r}_0$ is $\mathcal{C}_{4z}$/$\mathcal{C}_{6z}$ symmetric}\\
     \end{cases},
\end{align}
where $\delta(\mathbf{r}-\mathbf{r}_0)$ is the Dirac delta function, and we used $\tilde{c} = |\tilde{c}|e^{i\phi_0}$, $\tilde{c}_1 = |\tilde{c}|e^{i\phi_1}$, $z = x+iy= |z|e^{i\phi}$, and the fact that $\partial_z\overline{\partial_z}\mathcal{O}(\epsilon^2)$ is finite. This 
calculation shows that there is a singularity in the magnetic field distribution. Specifically, at the point where $\psi_\Gamma(\mathbf{r})$ has a zero there is a delta function magnetic field with flux $-\Phi_0 =  -2\pi$ (in the units of $\hbar/e$). Since, $K_\lambda(\mathbf{r})$ has a total of zero magnetic flux through the moir\'e unti cell, the total flux through moir\'e unit cell everywhere else due to $K_\lambda(\mathbf{r})$ is $\Phi_0 =  2\pi$. Furthermore, there are singular terms proportional to $1/|z|^2$ and $1/|z|$ in the $\mathcal{C}_{2z}$ and $\mathcal{C}_{3z}$ cases, however, since their integral over $\phi$ around $z=0$ are zero, so they do not contribute to the flux. These singularities in the magnetic field distribution for the flat bands in the single QBCP system is in contrast to the twisted bilayer graphene (TBG) case, where there is no singularity in the magnetic field distribution~\cite{ledwith2020fractional} (this is because in case of TBG, the wavefunction $\psi_K(\mathbf{r})$ near its zero behaves like $\psi_K(\mathbf{r}_0+\epsilon(x,y))\sim (x+iy)$ and hence $\frac{\psi_K(\mathbf{r}_0+\epsilon(x,y))}{\vartheta\left(\frac{x+iy}{a_1},\tau\right)}\sim \text{const.}+\mathcal{O}(\epsilon)$).

Another important point worth mentioning is that since $\vartheta\left(\frac{z-z_0}{a_1}-\frac{k}{b_2},\tau\right)$ has zeros at $\mathbf{r} = \mathbf{r}_0-\frac{1}{2\pi}(\hat{z}\cdot\mathbf{a}_1\times\mathbf{a}_2)(\hat{z}\times\mathbf{k})+m \mathbf{a}_1+n\mathbf{a}_2$, both moir\'e flatband wavefunction $\psi_\mathbf{k}(\mathbf{r})$ and LLL wavefunction $\psi_\mathbf{k}^\text{LLL}(\mathbf{r})$ have zeros at this momentum dependent position around which the wavefunctions behave as $\psi_\mathbf{k}(\mathbf{r}_0+\epsilon(x,y))\sim \epsilon(x+iy)$.

\subsection{Construction of wave-functions in the case of four flat bands (two flat bands per sublattice) and similarity to the lowest Landau level wavefunctions for a choice of unit cell with $4\pi$ flux}\label{sec:FourFlatBand}
In the case of four flat bands, we see in {\color{red}Fig.~1(a)} that the sublattice-polarized wavefunction $\psi_\Gamma(\mathbf{r})$ has two zeros at the two corners $\mathbf{r}_0^{(1)}$ and $\mathbf{r}_0^{(2)}$ of the moir\'e unit cell. So, two functions $f_\mathbf{k}(z;\mathbf{r}_0^{(1)})$ and $f_\mathbf{k}(z;\mathbf{r}_0^{(2)})$ can be created, and two sublattice-polarized flatband wavefunctions $\psi_\mathbf{k}^{(1)}(\mathbf{r}) = f_\mathbf{k}(z;\mathbf{r}_0^{(1)})\psi_\Gamma(\mathbf{r})$ and $\psi_\mathbf{k}^{(1)}(\mathbf{r}) = f_\mathbf{k}(z;\mathbf{r}_0^{(2)})\psi_\Gamma(\mathbf{r})$ can be otained. The other two flatband wavefunctions (polarized on the other sublattice) are related to these by time reversal symmetry. Indeed, in~\cite{sarkar2023symmetry}, this is the way flat band wavefunctions were created for multiple flatband per sublattice cases. It was further noticed that the two sublattice-polarized wave-functions $\psi_\mathbf{k}^{(1)}(\mathbf{r})$ and $\psi_\mathbf{k}^{(2)}(\mathbf{r})$ are independent for generic momenta $\mathbf{k}$, but they are not orthogonal. Also, since $f_{\mathbf{k}=\mathbf{0}}(z;\mathbf{r}_0^{(1)}) = f_{\mathbf{k}=\mathbf{0}}(z;\mathbf{r}_0^{(2)}) = 1$, at the $\Gamma$ point, there is a singularity in this construction; the second $\Gamma$ point wave-function (one of them $\psi_\Gamma(\mathbf{r})$ is already known) separately by performing an analytical continuation in $\mathbf{k}$ (see~\cite{sarkar2023symmetry} for more details). Moreover, this singularity at the $\Gamma$ point ($\mathbf{k} = \mathbf{0}$) gives rise a nontrivial Berry phase winding around $\mathbf{k} = \mathbf{0}$ in one of the wave-functions; this extra Berry phase winding cancels the Chern number of that wave-function resulting in the two bands together having total Chern number of $|C| =1$ (see~\cite{sarkar2023symmetry} for details). Lastly, it was proven using general form of these wavefunctions in~\cite{sarkar2023symmetry}, that two flatbands per sublattice together have ideal non-Abelian quantum geometry; $\text{tr}(g^{mn}_{\alpha \beta}(\mathbf{k})) = \sum_{n}'\sum_{\alpha \in \{x,y\}} g^{nn}_{\alpha\alpha}(\mathbf{k}) = |\text{tr}(F_{xy}^{mn}(\mathbf{k}))| = |\sum_{n}'F_{xy}^{nn}(\mathbf{k})|$, where $g_{\alpha\beta}^{mn}(\mathbf{k})$ is the non-Abelian quantum metric and $F_{xy}^{mn}(\mathbf{k})$ is the non-Abelian Berry curvature, and the sum is over the flatbands polarized on one sublattice. We numerically show this for the model in Fig.~1 of the main text in Fig.~\ref{fig:s6}(a-c).

Although the procedure mentioned above is a simple way to extend the flatband construction of wavefunction for single flatband per sublattice to the multiple flatband case, below we describe another (equivalent) construction of the wavefunctions for the multiple flatband per sublattice case that reveals a hidden structure. To start, recall that the LLL wavefunction has one zero per magnetic unit cell through which $2\pi$ (in the units of $\hbar/e$) flux is threaded. If we choose to double the magnetic unit cell such that the unit cell has $4\pi$ magnetic flux, the LLL wavefunction have two zeros in such a unit cell, and also doubling the unit cell halves the Brillouin zone folding the LLL band into two bands. Note further that the wavefunction $\psi_\Gamma(\mathbf{r})$ in the four flatband case has two zeros in the moir\'e unit cell. These two facts together allude that the wavefunctions of two moir\'e flatbands per sublattice (in the four flatband case) can be mapped to LLL wavefunctions for unit cells with $4\pi$ flux (similar to the mapping between wavefunction for single moir\'e flatband per sublattice and LLL wavefunctions for unit cells with $2\pi$ flux shown in the previous section).

Following the argument of the previous section, the two trial wavefunctions are still of the form $\psi_\mathbf{k}^{(i)}(\mathbf{r}) = f_\mathbf{k}^{(i)}(z)\psi_\Gamma(\mathbf{r})$ (for $i=1,2$), where the two functions $f_\mathbf{k}^{(1)}(z)$ and $f_\mathbf{k}^{(2)}(z)$ are still Bloch periodic holomorphic functions. We further require that $f_\mathbf{k}^{(1)}(z)$ and $f_\mathbf{k}^{(2)}(z)$ have poles at positions of the zeros of $\psi_\Gamma(\mathbf{r})$. Using the properties of Jacobi theta function, we see that the functions
\begin{equation}
\begin{split}
     f_{\mathbf{k};p}^{(1)}(z) &= e^{i (\mathbf{k}\cdot\mathbf{a}_1) z/a_1}\frac{\vartheta\left(\frac{z-z_0^{(1)}}{a_1}-p\frac{k}{b_2},\tau\right)\vartheta\left(\frac{z-z_0^{(2)}}{a_1}-(1-p)\frac{k}{b_2},\tau\right)}{\vartheta\left(\frac{z-z_0^{(1)}}{a_1},\tau\right)\vartheta\left(\frac{z-z_0^{(2)}}{a_1},\tau\right)},\\
     f_{\mathbf{k};p}^{(2)}(z) &= e^{i (\mathbf{k}\cdot\mathbf{a}_1) z/a_1}\frac{\vartheta\left(\frac{z-z_0^{(1)}}{a_1}-(1-p)\frac{k+b_2}{b_2},\tau\right)\vartheta\left(\frac{z-z_0^{(2)}}{a_1}-p\frac{k+b_2}{b_2},\tau\right)}{\vartheta\left(\frac{z-z_0^{(1)}}{a_1},\tau\right)\vartheta\left(\frac{z-z_0^{(2)}}{a_1},\tau\right)},
\end{split}
\end{equation}
are independent of each other for $0\leq p \leq 1$, and both satisfy Bloch periodicity, and have poles at $\mathbf{r}_0^{(1)}$ and $\mathbf{r}_0^{(2)}$, where $\psi_\Gamma(\mathbf{r})$ has zeros. Hence, the wavefunctions 
\begin{equation}
     \begin{split}
          \psi_{\mathbf{k};p}^{(1)}(\mathbf{r}) &= e^{i (\mathbf{k}\cdot\mathbf{a}_1) z/a_1}\vartheta\left(\frac{z-z_0^{(1)}}{a_1}-p\frac{k}{b_2},\tau\right)\vartheta\left(\frac{z-z_0^{(2)}}{a_1}-(1-p)\frac{k}{b_2},\tau\right)\frac{\psi_\Gamma(\mathbf{r})}{\vartheta\left(\frac{z-z_0^{(1)}}{a_1},\tau\right)\vartheta\left(\frac{z-z_0^{(2)}}{a_1},\tau\right)},\\
          \psi_{\mathbf{k};p}^{(2)}(\mathbf{r}) &= e^{i (\mathbf{k}\cdot\mathbf{a}_1) z/a_1}\vartheta\left(\frac{z-z_0^{(1)}}{a_1}-(1-p)\frac{k+b_2}{b_2},\tau\right)\vartheta\left(\frac{z-z_0^{(2)}}{a_1}-p\frac{k+b_2}{b_2},\tau\right)\frac{\psi_\Gamma(\mathbf{r})}{\vartheta\left(\frac{z-z_0^{(1)}}{a_1},\tau\right)\vartheta\left(\frac{z-z_0^{(2)}}{a_1},\tau\right)},
     \end{split}
\end{equation}
are smooth bounded Bloch-periodic functions. The wavefunctions for every such choice of $p$ ($0\leq p \leq 1$) are unitarily related to every other choice (this can be verified numerically). Notice further that if we choose $p=1$, the wavefunctions are exactly the same as the wavefunctions we discussed in the first paragraph of this subsection. Here we fix $p=1/2$ for a particular reason which we describe now. For $p=1/2$, we have
\begin{equation}\label{eq:doubleWFs}
     \begin{split}
          \psi_{\mathbf{k}}^{(1)}(\mathbf{r})\equiv\psi_{\mathbf{k};1/2}^{(1)}(\mathbf{r}) &= e^{i (\mathbf{k}\cdot\mathbf{a}_1) z/a_1}\vartheta\left(\frac{z-z_0^{(1)}}{a_1}-\frac{k}{2b_2},\tau\right)\vartheta\left(\frac{z-z_0^{(2)}}{a_1}-\frac{k}{2b_2},\tau\right)\frac{\psi_\Gamma(\mathbf{r})}{\vartheta\left(\frac{z-z_0^{(1)}}{a_1},\tau\right)\vartheta\left(\frac{z-z_0^{(2)}}{a_1},\tau\right)},\\
          \psi_{\mathbf{k}}^{(2)}(\mathbf{r})\equiv\psi_{\mathbf{k};1/2}^{(2)}(\mathbf{r}) &= e^{i (\mathbf{k}\cdot\mathbf{a}_1) z/a_1}\vartheta\left(\frac{z-z_0^{(1)}}{a_1}-\frac{k+b_2}{2b_2},\tau\right)\vartheta\left(\frac{z-z_0^{(2)}}{a_1}-\frac{k+b_2}{2b_2},\tau\right)\frac{\psi_\Gamma(\mathbf{r})}{\vartheta\left(\frac{z-z_0^{(1)}}{a_1},\tau\right)\vartheta\left(\frac{z-z_0^{(2)}}{a_1},\tau\right)}.
     \end{split}
\end{equation}
These two functions satisfy
\begin{equation}
     \psi_{\mathbf{k}+\mathbf{b}_2}^{(1)}(\mathbf{r}) = \psi_{\mathbf{k}}^{(2)}(\mathbf{r}),\; \psi_{\mathbf{k}+\mathbf{b}_2}^{(2)}(\mathbf{r}) = \psi_{\mathbf{k}}^{(1)}(\mathbf{r}) \text{, and } \psi_{\mathbf{k}+2\mathbf{b}_2}^{(i)}(\mathbf{r}) = \psi_{\mathbf{k}}^{(i)}(\mathbf{r})
\end{equation}
\textit{which seems to suggest that the moir\'e Brillouin zone (given by $\mathbf{b}_1$ and $\mathbf{b}_2$ defined earlier) can be unfolded to twice as large a moir\'e Brillouin zone given by $\mathbf{b}_1$ and $2\mathbf{b}_2$ with only one band per sublattice. If this were true, then the moir\'e unit cell could have been halved into a unit cell given by $\mathbf{a}_1$ and $\mathbf{a}_2/2$. Later, we are also going to see that the moir\'e unit cell given by $\mathbf{a}_1$ and $\mathbf{a}_2$ has $4\pi$ magnetic flux, which would also suggest that moir\'e unit cell could in principle be halved. However, crucially, the function $\psi_\Gamma(\mathbf{r})$ does not have the periodicity of $\mathbf{a}_2/2$, it has periodicity of $\mathbf{a}_2$. Hence, the moir\'e Brillouin zone cannot actually be unfolded to a larger Brillouin zone.}

Next, to show how these wavefunctions are analogous to LLL wavefunctions with $4\pi$ flux per unit cell, we again start from the Hamiltonian in Eq.~\eqref{eq:LLHam}, but this time we use a vector potential $\mathbf{A}$ that corresponds to $4\pi$ magnetic flux per unit cell. Specifically, we will use 
\begin{equation}
     \mathbf{A}(\mathbf{r}) = B_0(\hat{b}_2\cdot(\mathbf{r}-\mathbf{r}_0^{(1)}))(\hat{z}\times\hat{b}_2)+B_0(\hat{b}_2\cdot(\mathbf{r}-\mathbf{r}_0^{(2)}))(\hat{z}\times\hat{b}_2),
\end{equation}
where $B_0 (\hat{z}\cdot(\mathbf{a}_1\times\mathbf{a}_2)) = 2\pi \hbar/e$, $\hat{b}_2 = \mathbf{b}_2/|\mathbf{b}_2|$ is the unit vector along the reciprocal lattice vector $\mathbf{b}_2$, such that magnetic field corresponding to $\mathbf{A}(\mathbf{r})$ is $2B_0$, and flux per unit cell given by $\mathbf{a}_1$ and $\mathbf{a}_2$ is $4\pi\hbar/e$. Here $\mathbf{r}_0^{(1)}$ and $\mathbf{r}_0^{(2)}$ are the same positions where $\psi_\Gamma(\mathbf{r})$ of the moir\'e system has zeros. We made this choice for $\mathbf{A}(\mathbf{r})$ because it is going to place the zeros of the LLL wavefunctions at the same places as the ones in $\psi_\mathbf{k}^{(i)}(\mathbf{r})$ (note that we are free to choose the position of zeros of the LLL wavefunctions as long as there is one zero per $2\pi$ flux quantum). Plugging the ansatz for LL eigenstate as $\psi^\text{LLL}(\mathbf{r}) = g(\mathbf{r}) \exp\left(-\frac{1}{2\ell_B^2}\{(\hat{b}_2\cdot(\mathbf{r}-\mathbf{r}_0^{(1)}))^2+(\hat{b}_2\cdot(\mathbf{r}-\mathbf{r}_0^{(2)}))^2\}\right)$ (note that here $\ell_B^2 = \hbar/(eB_0)$ and not $\hbar/(2eB_0)$), we find
\begin{equation}
     \frac{\hbar^2}{2m}\left[4\partial_z+\sum_{i=1}^2\frac{4}{\ell_B^2}(\hat{b}_2\cdot(\mathbf{r}-\mathbf{r}_0^{(i)}))(\hat{b}_{2x}+i\hat{b}_{2y})\right]\overline{\partial_z} g(\mathbf{r}) = (E-\frac{1}{2}\hbar\omega_c)g(\mathbf{r}).
\end{equation}
If $\overline{\partial_z} g(\mathbf{r}) = 0$, $E= \frac{1}{2}\hbar\omega_c$ is the lowest Landau level energy. This implies for LLL wavefunctions, $g$ is a holomorphic function of $z$. Next, we require the magnetic flux through a magnetic unit cell (given by $\mathbf{a}_1$ and $\mathbf{a}_2$) is $4\pi$ in the units of $\hbar/e$ (two flux quanta per unit cell), or in other words $2\pi \ell_B^2 =\hbar/(eB_0) = \hat{z}\cdot(\mathbf{a}_1\times\mathbf{a}_2)$ (recall that the magnetic field is $2B_0$, which makes the flux $4\pi$ per unit cell). Since the new magnetic unit cell now has $4\pi$ flux, which is twice as large as the smallest magnetic unit cell, the new Brillouin zone has half the area of the standard Brillouin zone, and hence the LLL band gets folded into two bands. Therefore, we want to construct two wavefunctions $\psi^\text{LLL}_{\mathbf{k},1}(\mathbf{r})$ and $\psi^\text{LLL}_{\mathbf{k},2}(\mathbf{r})$ per momentum that satisfy magnetic Bloch-periodicity
\begin{equation}
     \psi^\text{LLL}_{\mathbf{k},i}(\mathbf{r}+\mathbf{a}_1) = e^{i\mathbf{k}\cdot\mathbf{a}_1}\psi^\text{LLL}_{\mathbf{k},i}(\mathbf{r}),\,\psi^\text{LLL}_{\mathbf{k},i}(\mathbf{r}+\mathbf{a}_2) = e^{i\mathbf{k}\cdot\mathbf{a}_2}e^{-2\pi i\frac{\mathbf{a}_1\cdot(\mathbf{r}-\mathbf{r}_0^{(1)}+\mathbf{a}_2)}{|\mathbf{a}_1|^2}}e^{-2\pi i\frac{\mathbf{a}_1\cdot(\mathbf{r}-\mathbf{r}_0^{(2)}+\mathbf{a}_2)}{|\mathbf{a}_1|^2}}\psi^\text{LLL}_{\mathbf{k},i}(\mathbf{r}).
\end{equation}
Using the properties of Jacobi theta function, two such functions can be constructed as
\begin{equation}\label{eq:LLLWF4pi}
     \begin{split}
          \psi_{\mathbf{k},1}^\text{LLL}(\mathbf{r})&= e^{i (\mathbf{k}\cdot\mathbf{a}_1) z/a_1}\vartheta\left(\frac{z-z_0^{(1)}}{a_1}-\frac{k}{2b_2},\tau\right)\vartheta\left(\frac{z-z_0^{(2)}}{a_1}-\frac{k}{2b_2},\tau\right)e^{-\frac{\pi}{\hat{z}\cdot(\mathbf{a}_1\times\mathbf{a}_2)}\{(\hat{b}_2\cdot(\mathbf{r}-\mathbf{r}_0^{(1)}))^2+(\hat{b}_2\cdot(\mathbf{r}-\mathbf{r}_0^{(2)}))^2\}},\\
          \psi_{\mathbf{k},2}^\text{LLL}(\mathbf{r})&= e^{i (\mathbf{k}\cdot\mathbf{a}_1) z/a_1}\vartheta\left(\frac{z-z_0^{(1)}}{a_1}-\frac{k+b_2}{2b_2},\tau\right)\vartheta\left(\frac{z-z_0^{(2)}}{a_1}-\frac{k+b_2}{2b_2},\tau\right)e^{-\frac{\pi}{\hat{z}\cdot(\mathbf{a}_1\times\mathbf{a}_2)}\{(\hat{b}_2\cdot(\mathbf{r}-\mathbf{r}_0^{(1)}))^2+(\hat{b}_2\cdot(\mathbf{r}-\mathbf{r}_0^{(2)}))^2\}}.
     \end{split}
\end{equation}
These expressions are very similar to the ones in Eq.~\eqref{eq:doubleWFs}; the only difference is in the $\mathbf{k}$ independent factors in the functions $\frac{\psi_\Gamma(\mathbf{r})}{\vartheta\left(\frac{z-z_0^{(1)}}{a_1},\tau\right)\vartheta\left(\frac{z-z_0^{(2)}}{a_1},\tau\right)}$ vs $e^{-\frac{\pi}{\hat{z}\cdot(\mathbf{a}_1\times\mathbf{a}_2)}\{(\hat{b}_2\cdot(\mathbf{r}-\mathbf{r}_0^{(1)}))^2+(\hat{b}_2\cdot(\mathbf{r}-\mathbf{r}_0^{(2)}))^2\}}$. Using this analogy between LLL wavefunction and the moir\'e flatband wavefunction, we can rewrite the moir\'e flatband wavefunctions as
\begin{equation}
     \begin{split}
          &\psi_\mathbf{k}^{(i)}(\mathbf{r})\\
          &= \psi^\text{LLL}_{\mathbf{k},i}(\mathbf{r})\frac{\psi_\Gamma(\mathbf{r})}{\psi^\text{LLL}_{\Gamma,1}(\mathbf{r})}\\
          &=\psi^\text{LLL}_{\mathbf{k},i}(\mathbf{r}) \left|\frac{\psi_\Gamma(\mathbf{r})}{\psi^\text{LLL}_{\Gamma,1}(\mathbf{r})}\right|\zeta(\mathbf{r})\\
          &=h_\mathbf{k}^{(i)}(z) e^{-K(\mathbf{r})/4\ell_B^2}\zeta(\mathbf{r}) =\phi_\mathbf{k}^{(i)}(\mathbf{r})\zeta(\mathbf{r}),
     \end{split}
\end{equation}
where 
\begin{equation}
\begin{split}
     \phi_\mathbf{k}^{(1)}(\mathbf{r})e^{K(\mathbf{r})/4\ell_B^2} &=h_\mathbf{k}^{(1)}(z) = e^{i (\mathbf{k}\cdot\mathbf{a}_1) z/a_1}\vartheta\left(\frac{z-z_0^{(1)}}{a_1}-\frac{k}{2b_2},\tau\right)\vartheta\left(\frac{z-z_0^{(2)}}{a_1}-\frac{k}{2b_2},\tau\right)\\
     \phi_\mathbf{k}^{(2)}(\mathbf{r})e^{K(\mathbf{r})/4\ell_B^2} &=h_\mathbf{k}^{(2)}(z) = e^{i (\mathbf{k}\cdot\mathbf{a}_1) z/a_1}\vartheta\left(\frac{z-z_0^{(1)}}{a_1}-\frac{k+b_2}{2b_2},\tau\right)\vartheta\left(\frac{z-z_0^{(2)}}{a_1}-\frac{k+b_2}{2b_2},\tau\right)
\end{split}
\end{equation}
are holomorphic functions of $z$,  and 
\begin{equation}
e^{-K(\mathbf{r})/4\ell_B^2} = \left|\frac{\psi_\Gamma(\mathbf{r})}{\vartheta\left(\frac{z-z_0^{(1)}}{a_1},\tau\right)\vartheta\left(\frac{z-z_0^{(2)}}{a_1},\tau\right)}\right|,\; \zeta(\mathbf{r}) = \exp\left(i\arg\left(\frac{\psi_\Gamma(\mathbf{r})}{\vartheta\left(\frac{z-z_0^{(1)}}{a_1},\tau\right)\vartheta\left(\frac{z-z_0^{(2)}}{a_1},\tau\right)}\right)\right).
\end{equation}
Analougous to the single flatband per sublattice case, here $\phi_\mathbf{k}^{(i)}(\mathbf{r}) =h_\mathbf{k}^{(i)}(z)e^{-K(\mathbf{r})/4\ell_B^2}$ have the interpretation of LLL wavefunctions in spatially varying magnetic field with $4\pi$ flux per unit cell; where the magnetic field distribution can be obtained using $B(\mathbf{r}) = B_0 \partial_z\overline{\partial_z}K(\mathbf{r})$. The LLL wavefunctions in Eq.~\eqref{eq:LLLWF4pi} have the form $ \psi^\text{LLL}_{\mathbf{k},i}(\mathbf{r})=h_\mathbf{k}^{(i)}(z)e^{-K_0(\mathbf{r})/4\ell_B^2}$ with $K_0(\mathbf{r}) = 2\{(\hat{b}_2\cdot(\mathbf{r}-\mathbf{r}_0^{(1)}))^2+(\hat{b}_2\cdot(\mathbf{r}-\mathbf{r}_0^{(2)}))^2\}$ and magnetic field $B(\mathbf{r}) = B_0 \partial_z\overline{\partial_z}K_0(\mathbf{r}) = 2B_0$. Again, in the case of moir\'e flatband wavefunction, the function $K(\mathbf{r})$ can be split into two parts
\begin{equation}
     K(\mathbf{r}) = K_0(\mathbf{r})+K_\lambda(\mathbf{r}), 
\end{equation}
where $K_0(\mathbf{r})$ acounts for the average magnetic field $2B_0$, and $K_\lambda(\mathbf{r})$ acounts for the spatial variation of the magnetic field about the mean. Explicitly written, 
\begin{equation}\label{eq:MagneticField4Pi}
     K_\lambda(\mathbf{r}) = 2\ell_B^2 \log \left|\frac{\psi^\text{LLL}_{\Gamma,1}(\mathbf{r})}{\psi_\Gamma(\mathbf{r})}\right|^2 = \frac{\hat{z}\cdot(\mathbf{a}_1\times\mathbf{a}_2)}{\pi}\log \left|\frac{\psi^\text{LLL}_{\Gamma,1}(\mathbf{r})}{\psi_\Gamma(\mathbf{r})}\right|^2, \text{ and } \frac{B(\mathbf{r})}{B_0} = 2+  \frac{\hat{z}\cdot(\mathbf{a}_1\times\mathbf{a}_2)}{\pi}\partial_z\overline{\partial_z}\log \left|\frac{\psi^\text{LLL}_{\Gamma,1}(\mathbf{r})}{\psi_\Gamma(\mathbf{r})}\right|^2.
\end{equation}
It is worth mentioning that since $\psi^\text{LLL}_{\Gamma,1}(\mathbf{r})$ satisfies magnetic translation symmetry and $\psi_\Gamma(\mathbf{r})$ satisfies ordinary translation symmetry, $K_\lambda(\mathbf{r})$ is a periodic function with moir\'e periodicity. Hence, the magnetic flux of $K_\lambda(\mathbf{r})$ is zero in a moir\'e unit cell, and total magnetic flux of $K(\mathbf{r})$ is $4\pi$ in the units of $\hbar/e$. 

\begin{figure}[ht]
     \centering
\includegraphics[width=\textwidth]{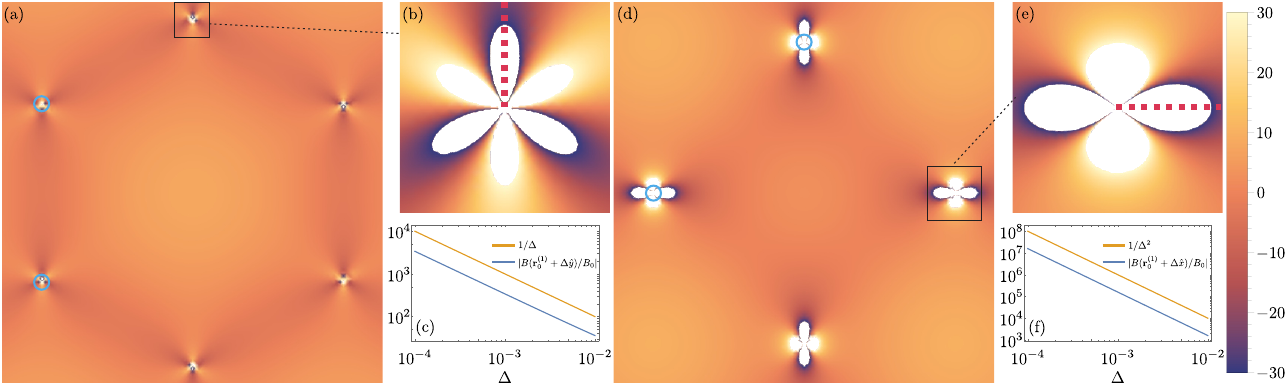}
     \caption{Magnetic field distributions in the moir\'e unit cell for the moir\'e systems considered in {\color{red}Fig.~1 and Fig.~3 of the main text}. (a) shows the magnetic field distribution for the moir\'e system in Fig.~1 of the main text. (b) shows the field distribution in a small region (shown by black square in (a)) near the corner of the unit cell, where there is a zero of $\psi_\Gamma(\mathbf{r})$. It shows a three fold symmetric distribution. (c) shows the magnetic field along the red dashed line as a function of distance from the corner of the unit cell. Clearly, the magnetic field distribution near the corner of the unit cell scales as $1/|\Delta|$, where $|\Delta|$ is the distance from the corner. This is in agreement with the prediction in Eq.~\eqref{eq:Bsingularity} for a $\mathcal{C}_{3z}$ symmetric point. (d) shows the magnetic field distribution for the moir\'e system in Fig.~4 of the main text for $\alpha=0.6651, \beta=0.7183$. (e) shows the field distribution in a small region (shown by black square in (d)) near the center of the edge of the unit cell, where there is a zero of $\psi_\Gamma(\mathbf{r})$. It shows a two fold symmetric distribution. (f) shows the magnetic field along the red dashed line as a function of distance from the center of the edge of the unit cell. Clearly, the magnetic field distribution near the center of the edge of the unit cell scales as $1/|\Delta|^2$, where $|\Delta|$ is the distance from the center of the edge. This is in agreement with the prediction in Eq.~\eqref{eq:Bsingularity} for a $\mathcal{C}_{2z}$ symmetric point. In (a) and (c) we show circular loops around the singularity points in blue, around which the numerical line integral of the vector potential corresponding to the magnetic field gives $-2\pi$ magnetic flux (in units of $\hbar/e$) confirming the existence of a Dirac delta function in the magnetic field distribution as predicted in Eq.~\eqref{eq:Bsingularity}.
    }
     \label{fig:s5}
\end{figure}

We plot the function $B(\mathbf{r})/B_0$ for the example system considered in {\color{red}Fig.~1} in {\color{red}Fig.~1(f)} and Fig.~\ref{fig:s5}(a-c). Importantly, since $\psi_\Gamma(\mathbf{r})$ has two zeros at the $\mathcal{C}_{3z}$ symmetric corners $\mathbf{r}_0^{(1)}$ and $\mathbf{r}_0^{(2)}$ of the moir\'e unit cell, the magnetic field $B(\mathbf{r})$ has a singular delta function contribution at $\mathbf{r}_0^{(i)}$ with corresponding magnetic flux $-2\pi$ and another singular $\frac{\cos(3\phi+\phi_1)}{|\mathbf{r}-\mathbf{r}_0^{(i)}|}$ (where $\phi$ is the angle of the vector $\mathbf{r}-\mathbf{r}_0^{(i)}$ from $x$ axis, and $\phi_1$ is  a constant) contribution that has zero flux at each zero as was shown in Eq.~\eqref{eq:Bsingularity}). The $\frac{\cos(3\phi+\phi_1)}{|\mathbf{r}-\mathbf{r}_0^{(i)}|}$ singularity at $\mathbf{r}_0^{(i)}$ can be seen in {\color{red}Fig.~\ref{fig:s5}(b-c)}. Since $K_\lambda(\mathbf{r})$ has total magnetic flux to be zero in the moir\'e unit cell, the flux of the magnetic field everywhere else in the moir\'e unit cell due to $K_\lambda(\mathbf{r})$ must be $4\pi$. We verified this numerically for the example of {\color{red}Fig.~1 of the main text}. 

We plot the function $B(\mathbf{r})/B_0$ for the example system considered in {\color{red}Fig.~3} in Fig.~\ref{fig:s5}(d-f). Importantly, since $\psi_\Gamma(\mathbf{r})$ has two zeros at the $\mathcal{C}_{2z}$ symmetric $2b$ Wyckoff positions $\mathbf{r}_0^{(1)}$ and $\mathbf{r}_0^{(2)}$ of the moir\'e unit cell (center of the edge of the moir\'e unit cell), the magnetic field $B(\mathbf{r})$ has a singular delta function contribution at $\mathbf{r}_0^{(i)}$ with corresponding magnetic flux $-2\pi$ and another singular $~\frac{1}{|\mathbf{r}-\mathbf{r}_0^{(i)}|^2}$ contribution that has zero flux at each zero as was shown in Eq.~\eqref{eq:Bsingularity}). The $\frac{1}{|\mathbf{r}-\mathbf{r}_0^{(i)}|^2}$ singularity at $\mathbf{r}_0^{(i)}$ can be seen in {\color{red}Fig.~\ref{fig:s5}(e-f)}. Since $K_\lambda(\mathbf{r})$ has total magnetic flux to be zero in the moir\'e unit cell, the flux of the magnetic field everywhere else in the moir\'e unit cell due to $K_\lambda(\mathbf{r})$ must be $4\pi$. We verified this numerically for the example of {\color{red}Fig.~3  of the main text}.

\section{Construction of one maximally localized Wannier function per unit cell from the two flat bands per sublattice}\label{sec:Wannierization}
As was discussed earlier, the two degenerate flatbands per sublattice have total Chern number $|C|=1$. Hence, it may be possible to gauge fix the two flatband wavefunctions such that one of them corresponds to a locallized Wannier orbital per unit cell. The example case in {\color{red}Fig.~1} of the main text has $p6mm$ wallpaper group symmetry, and the symmetry eigenvalues of the two bands at high symmetry points of the Brillouin zone are
\begin{equation}
\begin{split}
     &\Gamma \text{ point: } C_6 \text{ eigenvalues }\eta_i= e^{4\pi i/3},1,\\
     &M \text{ point: } C_2 \text{ eigenvalues }\theta_i= +1,-1,\\
     &K \text{ point: } C_3 \text{ eigenvalues }\zeta_i= e^{2\pi i/3},e^{4\pi i/3}.\\
\end{split}
\end{equation}
Hence, using Fang-Gilbert-Bernevig formula~\cite{fang2012bulk} $e^{2\pi i C/6} = \prod_{i\in \text{bands}}\eta_i\theta_i\zeta_i$, we find $e^{2\pi i C/6} = e^{\pi i/3}$ in the case of the example case in {\color{red}Fig.~1}, consistent with total Chern number of the two bands being $|C|=1$. Now, if we could split the two bands (fix the gauge of the wavefunction of each band at every $\mathbf{k}$ point smoothly), such that one of the bands has symmetry eigen values $e^{4\pi i/3},1,e^{2\pi i/3}$ at $\Gamma$, $M$ and $K$ points, respectevily, that band would be topologically trivial according to Fang-Gilbert-Bernevig formula. Furthermore, such a band corresponds to a localized Wannier function at the $1a$ Wyckoff position of the moir\'e unit cell and transforms under $\mathcal{C}_{6z}$ as $^1E_2$ or in other words like a $d_{x^2-y^2}+id_{2xy}$ orbital according to Bilbao Crystallography Server~\cite{aroyo2011crystallographys,aroyo2006bilbaoI,aroyo2006bilbaoII}. In the following we show the numerical gauge fixing procedure using Wannier90~\cite{marzari1997maximally,souza2001maximally,pizzi2020wannier90} that achieves this splitting of the two bands as shown in {\color{red}Fig.~1(b)}.

Let the basis states in the real space for the Hamiltonian in Eq.~eqref{eq:hamiltonian1} be $|\mathbf{r},A\rangle$ and $|\mathbf{r},B\rangle$. Fourier transform of these states gives
\begin{equation}
	|\mathbf{r}, \alpha\rangle = \sum_\mathbf{k}e^{-i\mathbf{k}\cdot\mathbf{r}}|\mathbf{k}, \alpha\rangle =  \sum_{\mathbf{k}\in \text{mBZ}}\sum_{\mathbf{b}}e^{-i(\mathbf{k}+\mathbf{b})\cdot\mathbf{r}}|\mathbf{k},\mathbf{b}, \alpha\rangle, \text{, }\alpha \in\{A,B\},
\end{equation}
where we broke down the sum over the $\mathbf{k}$ into sum over $\mathbf{k}$ in the BZ and sum over moir\'e reciprocal lattice vectors and $|\mathbf{k},\mathbf{b}, \alpha\rangle \equiv |\mathbf{k}+\mathbf{b}, \alpha\rangle$. Recall that these basis states have the following transformation properties
\begin{equation}
\begin{split}
	\mathcal{C}_{3z}\{|\mathbf{r}, A\rangle,|\mathbf{r}, B\rangle\} &= \{|\mathcal{C}_{3z}\mathbf{r}, A\rangle,|\mathcal{C}_{3z}\mathbf{r}, B\rangle\}\rho(\mathcal{C}_{3z}),\, \rho(\mathcal{C}_{3z}) = \text{Diag}\{e^{4\pi i/3},e^{2\pi i/3}\},\\
	\mathcal{C}_{2z}\{|\mathbf{r}, A\rangle,|\mathbf{r}, B\rangle\} &= \{|\mathcal{C}_{2z}\mathbf{r}, A\rangle,|\mathcal{C}_{2z}\mathbf{r}, B\rangle\}\rho(\mathcal{C}_{2z}),\, \rho(\mathcal{C}_{2z}) = \mathds{1},\\
	\mathcal{M}_{x}\{|\mathbf{r}, A\rangle,|\mathbf{r}, B\rangle\} &= \{|\mathcal{M}_{x}\mathbf{r}, A\rangle,|\mathcal{M}_{x}\mathbf{r}, B\rangle\}\rho(\mathcal{M}_{x}),\, \rho(\mathcal{M}_{x}) = \sigma_x,\\
	\mathcal{T}\{|\mathbf{r}, A\rangle,|\mathbf{r}, B\rangle\} &= \{|\mathbf{r}, A\rangle,|\mathbf{r}, B\rangle\}\rho(\mathcal{M}_{x}),\, \rho(\mathcal{T}) = \sigma_x,\\
	T_\mathbf{R}\{|\mathbf{r}, A\rangle,|\mathbf{r}, B\rangle\} &= \{|\mathbf{r}+\mathbf{R}, A\rangle,|\mathbf{r}+\mathbf{R}, B\rangle\}
\end{split}
\end{equation}
where we chose $\rho(\mathcal{C}_{2z}) = \mathds{1}$ to specify that the irrep label of the QBCP is $\Gamma_5$ in the notation of BCS (we could have just as easily chose $\rho(\mathcal{C}_{2z}) = -\mathds{1}$, then irrep would have been $\Gamma_6$). Also, here, $\mathbf{R} = n_1\mathbf{a}_1^m+n_2\mathbf{a}_2^m$ is a moir\'e lattice vector. 

A trial Wannier functions that is supported entirely on the $A$ sublattice and transforms as $d_{x^2-y^2}+id_{2xy}$ orbital at $1a$ Wyckoff position or the center of the Wigner-Seitz unit cell is
\begin{equation}
\begin{split}
	|W'_{\mathbf{R},d_{x^2-y^2}+id_{2xy}}\rangle &=\frac{1}{\Omega\sqrt{2\pi\lambda_0^2}} \int d^2\mathbf{r} e^{-(\mathbf{r}-\mathbf{R})^2/2\lambda_0^2} |\mathbf{r}, A\rangle = \frac{1}{{\color{black}\Omega}\sqrt{2\pi\lambda_0^2}} \int d^2\mathbf{r} e^{-(\mathbf{r}-\mathbf{R})^2/2\lambda_0^2} \sum_{\mathbf{k}\in \text{mBZ}}\sum_{\mathbf{b}}e^{-i(\mathbf{k}+\mathbf{b})\cdot\mathbf{r}}|\mathbf{k},\mathbf{b}, A\rangle\\
	&=\frac{\sqrt{2\pi\lambda_0^2}}{\Omega} \sum_{\mathbf{k}\in \text{mBZ}}\sum_{\mathbf{b}} e^{-i\mathbf{k}\cdot\mathbf{R}-\frac{1}{2}\lambda_0^2(\mathbf{k}+\mathbf{b})^2}|\mathbf{k},\mathbf{b}, A\rangle,
\end{split}
\end{equation}
(where $\Omega$ is the area of the moir\'e system, and the parameter $\lambda_0$ encodes the spread of the Wannier function) because the basis states $|\mathbf{r},\alpha\rangle$ transform as $\Gamma_5$ rep, which are also $d$-type. This is similar to the construction of $p_x\pm ip_y$ orbitals for the topological heavy fermion model of TBG~\cite{song2022magic}. However, constructing the other 4 Wannier functions is new in this system compared to TBG.  

\begin{figure}[ht]
     \centering
\includegraphics[width=\textwidth]{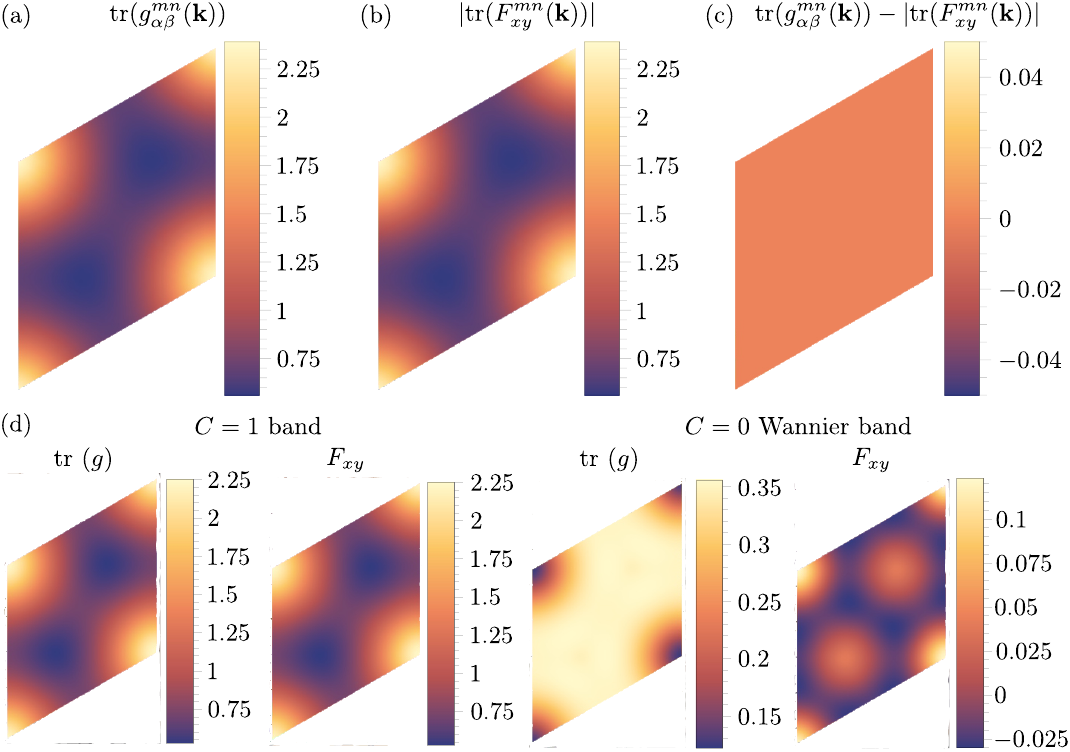}
     \caption{
     Quantum metric and Berry curvature of the two flatbands per sublattice in {\color{red}Fig.~1 of the main text}. (a) Trace $\text{tr}(g^{mn}_{\alpha \beta}(\mathbf{k})) = \sum_{n=1}^2\sum_{\alpha \in \{x,y\}} g^{nn}_{\alpha\alpha}(\mathbf{k})$ of the non-Abelian quantum metric $g_{\alpha\beta}^{mn} = \Re\left[\langle \partial_{k_\alpha}\tilde{u}_{N,\mathbf{k}}^{(m)}| \left(\mathds{1} - \sum_{n_1=1}^2 |\tilde{u}_{N,\mathbf{k}}^{(n_1)}\rangle\langle \tilde{u}_{N,\mathbf{k}}^{(n_1)}|\right)| \partial_{k_\beta}\tilde{u}_{N,\mathbf{k}}^{(n)}\rangle\right]$ of the two flatbands per sublattice, where $|\tilde{u}_{N,\mathbf{k}}^{(m)}\rangle$ ($m=1,2$) are the Bloch periodic parts of the wavefunctions $\psi_\mathbf{k}^{(m)}(\mathbf{r})$ after orthonormalization. (b) Trace $\text{tr}(F_{xy}^{mn}(\mathbf{k})) = \sum_{n=1}^2 F_{xy}^{nn}(\mathbf{k})$ of the non-Abelian Berry curvature $F_{xy}^{mn}(\mathbf{k}) = i \left(\langle\partial_{k_x}\tilde{u}_{N,\mathbf{k}}^{(m)}| \partial_{k_y}\tilde{u}_{N,\mathbf{k}}^{(n)}\rangle- \langle\partial_{k_y}\tilde{u}_{{N,\mathbf{k}}}^{(m)}| \partial_{k_x}\tilde{u}_{{N,\mathbf{k}}}^{(n)}\rangle\right)$ of the two flatbands per sublattice. (c) The difference $\text{tr}(g^{mn}_{\alpha \beta}(\mathbf{k})) - |\text{tr}(F_{xy}^{mn}(\mathbf{k}))|$ shows ideal non-Abelian quantum geometry of the two flatbands per sublattice. (d) the trace of Abelian quantum metric and the Berry curvature for the $C=1$ band and the Wannier band (the wavefunctions of these two bands were obtained using the procedure described in Sec.~\ref{sec:Wannierization}). The Chern band has ideal quantum geometry, but the Wannier band does not.
    }
     \label{fig:s6}
\end{figure}

Next we calculate the overlap between these trial Wannier functions and the energy eigenstates corresponding to the sublattice-polarized bands. Denoting the numerically obtained energy eigenstates as $|\psi_{\mathbf{k}^{(i)}}\rangle$ ($n=1,2$ for the two sublattice-polarized flatbands), we define the overlap matrix as
\begin{align}\label{eq:Overlap}
	A_{i}(\mathbf{k}) &\equiv \langle \psi_{\mathbf{k}}^{(i)}|W'_{\mathbf{0},d_{x^2-y^2}+id_{2xy}}\rangle =\frac{\sqrt{2\pi\lambda_0^2}}{\Omega} \sum_{\mathbf{k}\in \text{mBZ}}\sum_{\mathbf{b}} e^{-\frac{1}{2}\lambda_0^2(\mathbf{k}+\mathbf{b})^2}\langle \psi_{\mathbf{k}}^{(i)}|\mathbf{k},\mathbf{b}, A\rangle,
\end{align}
We feed the overlap matrix $A_{i}(\mathbf{k})$ ($n=1,2$) into the machinary of Wannier90~\cite{marzari1997maximally,souza2001maximally,pizzi2020wannier90} to construct the Maximally localized Wannier functions (MLWFs). We chose $\lambda_0 = a/10$ ($a$ is moir\'e lattice constant) for the numerical calculation, and used a $36\times 36$ grid to discretize the mBZ. Wannier90 returns MLWF in the plane wave basis $|\mathbf{k},\mathbf{b},\beta\rangle$ ($\beta = A,B$) as
\begin{equation}
	|W_{\mathbf{R},d_{x^2-y^2}+id_{2xy}}\rangle =\frac{1}{\Omega\sqrt{N}} \sum_{\mathbf{k}\in \text{BZ}}\sum_{\mathbf{b}} |\mathbf{k},\mathbf{b},\beta\rangle e^{-i\mathbf{k}\cdot\mathbf{R}} \tilde{v}_{\mathbf{b}\beta}(\mathbf{k}),
\end{equation}
where $N$ is the number of moir\'e unit cell. We can write the MLWFs in the real space basis
\begin{equation}
	w_{\beta}(\mathbf{r}-\mathbf{R})= \langle\mathbf{r},\beta|W_{\mathbf{R},\alpha}\rangle = \frac{1}{\Omega\sqrt{N}} \sum_{\mathbf{k}\in \text{mBZ}}\sum_{\mathbf{b}}  e^{i(\mathbf{k}+\mathbf{b})\cdot(\mathbf{r}-\mathbf{R})} \tilde{v}_{\mathbf{b}\beta,\alpha}(\mathbf{k}).
\end{equation}
The density plots of $\sum_\beta|w_{\beta}|^2$ are shown in {\color{red}Fig.~1(d)}. This Wannier function is supported entirely on the $A$ sublattice. The spread of this Wannier function is $\sqrt{\int d^2\mathbf{r} |\mathbf{r}|^2 |w_{\beta}(\mathbf{r})|^2/\int d^2\mathbf{r}  |w_{\beta}(\mathbf{r})|^2} \approx 0.21 a$, which means the Wannier function is really localized. Note that the wavefunction of this Wannier band at every $\mathbf{k}$ point is just the column vector $|\tilde{v}(\mathbf{k})\rangle$ whose rows are $\tilde{v}_{\mathbf{b}\beta}(\mathbf{k})$. From this we can obtain the wavefunction of the Chern band at every $\mathbf{k}$ point as eigenvectors of $(\sum_{n=1}^2|\psi_{\mathbf{k},n}\rangle\langle\psi_{\mathbf{k},n}|)-|\tilde{v}(\mathbf{k})\rangle\langle\tilde{v}(\mathbf{k})|$. The Wilson loop spectrum, the distribution of trace of the quantum metric in the BZ, and Berry curvature distribution in the BZ for both the Wannier band and Chern band are shown in {\color{red}Fig.~\ref{fig:s6}(d)}.
\newpage
\clearpage
\newpage
\section{$\sqrt{3}\times\sqrt{3}$ charge density wave state at $\nu=1/3$ filling for the model described in Fig.~1 of the main text}
\begin{figure}[ht]
     \centering
\includegraphics[width=\textwidth]{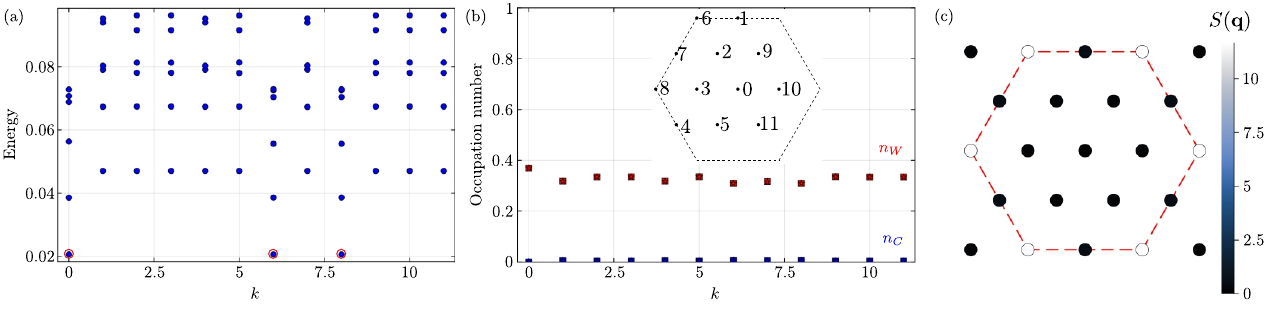}
     \caption{Charge density wave state at $\nu=1/3$ filling for the model described in Fig.~1 of the main text. (a) shows the energy spectrum obtained from exact diagonalization. The three quasi-degenerate ground states are circled by red circles. Their center-of-mass momentum are consistent with both a $\sqrt{3}\times\sqrt{3}$ CDW and a $\nu=1/3$ FCI state. (b) The occupation numbers of the Wannier band ($n_W$) and the Chern band ($n_C$) for the three ground states shown in red and blue, respectively. The circle, square, star shapes correspond to the ground states at momentum points $0$ (zone center), $6$ and $8$ (the two corners of the Brillouin zone), respectively. The occupation number distribution clearly shows that the Chern band is empty. Consistent with this observation, we also find the many body Chern number of the ground states to be zero. (c) Density-density correlation $S(\mathbf{q}) = \langle \rho(\mathbf{q})\rho(-\mathbf{q})\rangle - \langle \rho(\mathbf{q})\rangle\langle \rho(-\mathbf{q})\rangle$ (where $\rho(\mathbf{q})$ is density operator projected to the two sublattice-polarized bands) of the ground state showing clear peaks at the zone corners consistent with a $\sqrt{3}\times\sqrt{3}$ CDW state.}
     \label{fig:s3}
\end{figure}
\section{$\sqrt{3}\times\sqrt{3}$ CDW to FCI with many-body Chern number $C_\text{mb} = 1/3$ transition at $\nu=2/3$ filling for the model described in Fig.~1 of the main text by varying the inter-band tunneling} 
\begin{figure}[ht]
     \centering
\includegraphics[width=\textwidth]{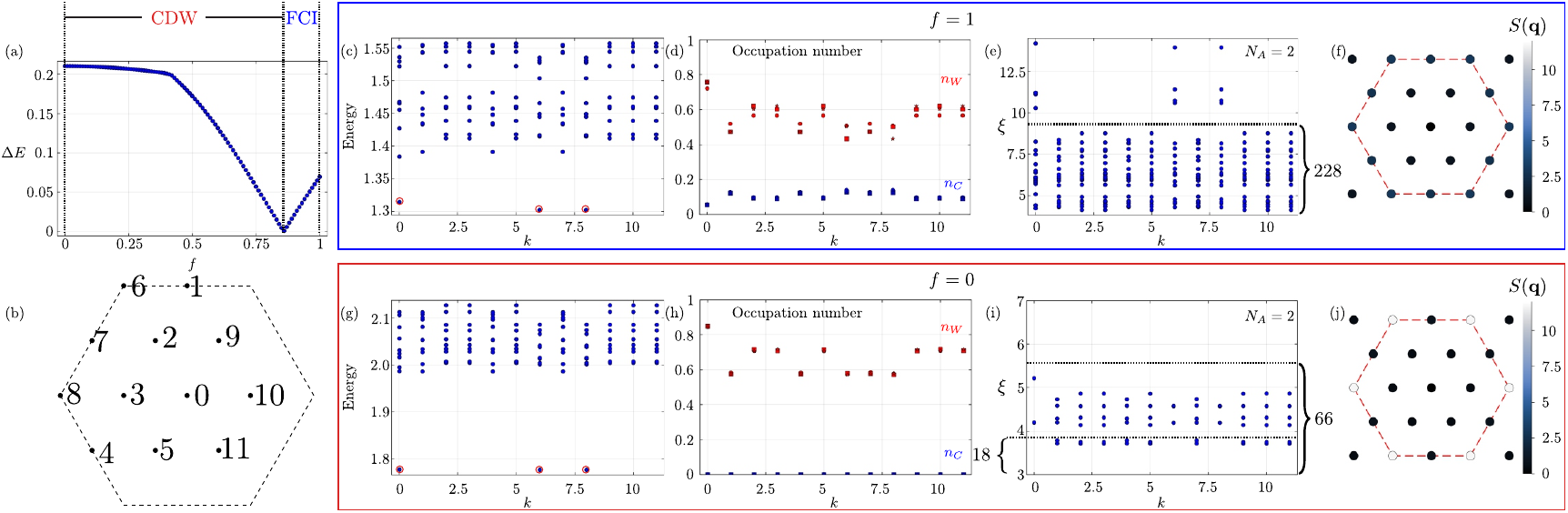}
     \caption{Transition between a $\sqrt{3}\times\sqrt{3}$ CDW to a FCI state with many-body Chern number $C_\text{mb} = 1/3$ by tuning the inter-band tunneling parameter $f$ at filling fraction $\nu=\frac{2}{3}$. (a) shows the many-body gap (the energy difference between the lowest energy excited state and the highest energy state among the three quasi-degenerate ground state) as function of the inter-band tunneling parameter $f$ defined in Eq.~\eqref{eq:f}. The gap closing point is $f\approx 0.89$. (b) shows momentum points and their numbering for the symmetric cluster used for the exact diagonalization. (c-f) Energy spectrum, occupation number of the two bands for the three quasi degenerate ground states, PES, and density-density correlation $S(\mathbf{q}) = \langle \rho(\mathbf{q})\rho(-\mathbf{q})\rangle - \langle \rho(\mathbf{q})\rangle\langle \rho(-\mathbf{q})\rangle$ for tunneling parameter $f=1$. (g-j) Energy spectrum, occupation number of the two bands for the three quasi degenerate ground states, PES, and density-density correlation $S(\mathbf{q}) = \langle \rho(\mathbf{q})\rho(-\mathbf{q})\rangle - \langle \rho(\mathbf{q})\rangle\langle \rho(-\mathbf{q})\rangle$ for tunneling parameter $f=0$. In (c) and (g), the quasi-degnerate ground states are encircled with red circles. In occupation number plots in (d) and (h) the occupation numbers of the Wannier band ($n_W$) and the Chern band ($n_C$) for the three ground states are shown in red and blue, respectively. The circle, square, star markers correspond to the ground states at momentum points $0$ (zone center), $6$ and $8$ (the two corners of the Brillouin zone), respectively. The PESs in (e) and (i) are calculated by separating the electrons into two groups with $N_A=2$ and $N_B= 6$ particles.
    }
     \label{fig:s1}
\end{figure}
In {\color{red}Fig.~2} of the main text we showed that the main-body ground state at $\nu=2/3$ filling for the model described in Fig.~1 of the main text is an FCI state with many-body Chern number $C_\text{mb} = 1/3$. Here we show that interband tunneling term in the projected density operator plays a crucial role in stabilizing this FCI state. Specifically, recall that the projected interaction is given by 
\begin{equation}
     H_\text{int} = \frac{1}{2A} \sum_\mathbf{q} V(\mathbf{q})\rho(\mathbf{q})\rho(-\mathbf{q}),
\end{equation}
where $A$ is the volume of the Brillouin zone, $V(\mathbf{q}) = 4\pi U \tanh(dq)/(\sqrt{3}qa)$ is the screen Coulomb interaction, $d$ is the separation between the two electrodes, $a$ is the moir\'e lattice constant, and we introduced the projected density operator 
\begin{equation}
    \rho(\mathbf{q}) = \sum_{\mathbf{k}}\sum_{i,j\in{C,W}}\lambda_{i,j,\mathbf{q}}(\mathbf{k})c^\dagger_{\mathbf{k},i}c_{\mathbf{k}+\mathbf{q},j},
\end{equation}
where $\lambda_{i,j,\mathbf{q}}(\mathbf{k}) = \langle \psi_\mathbf{k}^{(i)}(\mathbf{r})|e^{i\mathbf{q}\cdot\mathbf{r}}|\psi_{\mathbf{k}+\mathbf{q}}^{(j)}(\mathbf{r})\rangle$ is the form factor, and the indices $i,j$ go over the Chern band ($C$) and the Wannier band ($W$). We can write the band projected density explicitly as
\begin{equation}
\begin{split}
    \rho(\mathbf{q}) &= \sum_\mathbf{k} \lambda_{C,C,\mathbf{q}}(\mathbf{k})c^\dagger_{\mathbf{k},C}c_{\mathbf{k}+\mathbf{q},C} + \sum_\mathbf{k} \lambda_{W,W,\mathbf{q}}(\mathbf{k})c^\dagger_{\mathbf{k},W}c_{\mathbf{k}+\mathbf{q},W} + \sum_\mathbf{k} (\lambda_{C,W,\mathbf{q}}(\mathbf{k})c^\dagger_{\mathbf{k},C}c_{\mathbf{k}+\mathbf{q},W}+\lambda_{W,C,\mathbf{q}}(\mathbf{k})c^\dagger_{\mathbf{k},W}c_{\mathbf{k}+\mathbf{q},C})\\
    &\equiv\rho_{CC}(\mathbf{q})+\rho_{WW}(\mathbf{q})+\rho_{CW}(\mathbf{q}).
\end{split}
\end{equation}
The first two terms in this expression are the densities in the Chern band and the Wannier band, respectively. The last term is the tunneling term between the two bands. In this section we show that this last term is crucial for stabilizing the FCI state. To do that we artificially modify the density operator as
\begin{equation}\label{eq:f}
    \rho_f(\mathbf{q}) = \rho_{CC}(\mathbf{q})+\rho_{WW}(\mathbf{q})+f\rho_{CW}(\mathbf{q}),
\end{equation}
where we multiplied a factor $f$ to the tunneling term so we can vary it continuously. When $f = 0$, there is no tunneling between the two bands, whereas $f=1$ gives back the original full density operator. With this modified density operator, the new interacting Hamiltonian becomes
\begin{equation}
    H_{\text{int},f} = \frac{1}{2A} \sum_\mathbf{q} V(\mathbf{q})\rho_f(\mathbf{q})\rho_f(-\mathbf{q}).
\end{equation}

Results obtained by diagonalization this Hamiltonian at $f=0$ is shown in {\color{red}Figs.~\ref{fig:s1}(g-j)}. There are three quasi-degenerate ground states ({\color{red}Fig.~\ref{fig:s1}(g)}) at the center-of-mass momenta $0$ (zone center), $6$ and $8$ (zone corners), which is consistent with both an FCI state and $\sqrt{3}\times\sqrt{3}$ CDW state. However, the occupation number plot ({\color{red}Fig.~\ref{fig:s1}(h)}) shows that the Chern band is completely empty for $f=0$. Consistent with this, we verified that the many body Chern numbers of these ground states is $C_\text{mb} = 0$. Furthermore, the number of low-lying states in particle entangle spectrum in {\color{red}Fig.~\ref{fig:s1}(i)} is consistent with a $\nu=2/3$ CDW state as we show in Sec.~\ref{sec:PES}. The density-density correlation $S(\mathbf{q}) = \langle \rho(\mathbf{q})\rho(-\mathbf{q})\rangle - \langle \rho(\mathbf{q})\rangle\langle \rho(-\mathbf{q})\rangle$ in {\color{red}Fig.~\ref{fig:s1}(j)} shows pronounced peaks at zone corners further supporting the claim of a $\sqrt{3}\times\sqrt{3}$ CDW state.

We plot the many-body gap $\Delta E$ (the energy difference between the lowest energy excited state and the highest energy state among the three quasi-degenerate ground state) as a function of $f$ in {\color{red}Fig.~\ref{fig:s1}(a)}, which clearly shows a gap closing at $f\approx0.89$. For $f< 0.89$, the ground state is a $\sqrt{3}\times\sqrt{3}$ CDW state, whereas for $f> 0.89$, the ground state is an FCI state with many-body Chern number $C_\text{mb}=1/3$.

We also show the energy spectrum, occupation number of the two bands for the ground states, PES and density-density correlation function for $f=1$ in {\color{red}Figs.~\ref{fig:s1}(c-f)}, respectively. The number of states (228) below the gap in PES is consistent with $\nu=1/3$ FCI state in a $C=1$ Chern band with $N=24$ single particle states. We verified that the ground states in this case have many-body Chern number $C_\text{mb}=1/3$. The density-density correlation function is much more uniform than the CDW state at $f=0$ (no pronounced peaks at the zone corners).

\section{Single particle Hamiltonian and four flat bands (two per sublattice) in moir\'e system with $p4mm$ symmetry}
For our square lattice model with four flat bands (two per sublattice), we choose the following continuum Hamiltonian with a QBCP at $\Gamma$ point and a $p4mm$ symmetric moir\'e periodic strain field
\begin{equation}\label{eq:HSinglep4mm}
\begin{split}
    \mathcal{H}(\mathbf{r})
    &= \begin{pmatrix}\mathbf{0} &\text{h.c.}\\  \left(2i\overline{\partial_{z}}\right)^2 +\tilde{A}(\mathbf{r}) & \mathbf{0}\end{pmatrix}=\begin{pmatrix}\mathbf{0} &\text{h.c.}\\  \left(2i\overline{\partial_{z}}\right)^2 +\alpha\tilde{A}_{\tilde{\beta}}(\mathbf{r}) & \mathbf{0}\end{pmatrix},\\ 
    \tilde{A}_{\tilde{\beta}}(\mathbf{r}) &= -2i \left(\cos\left(\tilde{\mathbf{b}}_1\cdot \mathbf{r} \right)-\cos\left(\tilde{\mathbf{b}}_2 \cdot \mathbf{r} \right)\right)+2\tilde{\beta}\left(\cos\left(\mathbf{b}_1 \cdot \mathbf{r} \right)-\cos\left(\mathbf{b}_2\cdot \mathbf{r} \right)\right),\\
    \Rightarrow \tilde{A}(\mathbf{r}) \equiv\alpha\tilde{A}_{\tilde{\beta}}(\mathbf{r}) &=-2i\alpha \left(\cos\left(\tilde{\mathbf{b}}_1\cdot \mathbf{r} \right)-\cos\left(\tilde{\mathbf{b}}_2 \cdot \mathbf{r} \right)\right)+2\alpha\tilde{\beta}\left(\cos\left(\mathbf{b}_1 \cdot \mathbf{r} \right)-\cos\left(\mathbf{b}_2\cdot \mathbf{r} \right)\right)\\
     &=-2i\alpha \left(\cos\left(\tilde{\mathbf{b}}_1\cdot \mathbf{r} \right)-\cos\left(\tilde{\mathbf{b}}_2 \cdot \mathbf{r} \right)\right)+2\beta\left(\cos\left(\mathbf{b}_1 \cdot \mathbf{r} \right)-\cos\left(\mathbf{b}_2\cdot \mathbf{r} \right)\right),
\end{split}
\end{equation}
where $\beta = \tilde{\beta}\alpha$ and our choice of the Brillouin zone is spanned by $\mathbf{b}_1$ and $\mathbf{b}_2$, as shown in red in Fig.~\ref{fig:s2}(a). The reason for choosing the moir\'e potential is the following. When $\beta = 0$, the Brillouin zone can be unfolded to a twice as large Brillouin zone spanned by $\tilde{\mathbf{b}}_1$ and $\tilde{\mathbf{b}}_2$, as shown in blue in Fig.~\ref{fig:s2}(a). To find if this model support exact flat bands at certain ``magic'' values of $\alpha$, we utilize a method introduced in Ref.~\cite{becker2022mathematics}. 
Here we construct the Birman-Schwinger operator~\cite{becker2022mathematics, becker2021spectral}
\begin{equation}
    T_{\tilde{\beta}}(\mathbf{k};\mathbf{r}) = - (2i\overline{\partial_z}-k)^{-2}\tilde{A}_{\tilde{\beta}}(\mathbf{r}).
\end{equation}
where $k=k_x+ik_y$ is an arbitrary wavevector, and for given value of $\tilde{\beta} =\beta/\alpha$, compute the eigenvalues of this operator $\eta_\mathbf{k}$. If an $\eta_\mathbf{k}$ is independent of $\mathbf{k}$, it provides a ``magic" value of $\alpha=1/\eta_\mathbf{k}$,  at which one exact flat band per sublattice emerge. Now, if an eigenvalue $\eta_\mathbf{k}$ of $T_{\tilde{\beta}}(\mathbf{k};\mathbf{r})$ is doubly degenerate, then at $\alpha=1/\eta_\mathbf{k}$ and $\beta = \tilde{\beta}/\eta_\mathbf{k}$, we get four flat bands (two per sublattice). By numerically sweeping the region $-2\leq \tilde{\beta}\leq2$, and finding doubly degenerate eigenvalues of $T_{\tilde{\beta}}(\mathbf{k};\mathbf{r})$, we find the red line in the $(\alpha,\beta)$ plane shown in Fig.~\ref{fig:s2}(a) and {\color{red}Fig.~3} of the main text.
The band strcutures at two points on this line are shown in Figs.~\ref{fig:s2}(b,c). In both cases, we see four degenerate exact flat bands at $E = 0$. We show in Figs.~\ref{fig:s2}(b,c) that the wavefunction $\psi_\Gamma(\mathbf{r})$ has two zeros at the $2b$ Wyckoff positions (centers of edges) of the unit cell  (this unit cell is shown in red in Fig.~\ref{fig:s2}(a)) edges. Note that the two $2b$ Wyckoff positions are related to each other by $\mathcal{C}_{4z}$ symmetry, so if there is a zero at one of the $2b$ Wyckoff positions, then there must be a zero at the other $2b$ Wyckoff position. When $\beta=0$, these zeros are actually at the corner of the primitive unit cell (the primitive unit cell is shown in blue in Fig.~\ref{fig:s2}(a)). These two zeros allows for the construction of two exact flat bands per sublattice as described in Sec.~\ref{sec:FourFlatBand}. It was shown in~\cite{sarkar2023symmetry} that the number of parameters that must be tuned (co-dimension) to obtain zeros at $\mathcal{C}_{2v}$ symmetric $2b$ Wyckoff position of a Hamiltonian of the form~\eqref{eq:HSinglep4mm} with a $p4mm$ symmetric moir\'e strain field is 1. This is why in the 2-parameter plane $\alpha-\beta$, we find a line where the four exact flatbands (two per sublattice) occur. We show in {\color{red}Fig.~\ref{fig:s5}(d-f)} the magnetic field distribution in the moir\'e unit cell obtained using Eq.~\eqref{eq:MagneticField4Pi}. We numerically verify that the total magnetic flux per unit cell is $4\pi$ in the units of $\hbar/e$. Note that for $\beta\neq 0$, the the positions of the zeros of $\psi_\Gamma(\mathbf{r})$ are $\mathcal{C}_{2v}$ symmetric. Hence, according to Eq.~\eqref{eq:Bsingularity}, the magnetic field distribution has a Dirac delta function singularity and a $1/|\mathbf{r}-\mathbf{r}_0^{(i)}|^2$ type singularity at each of the zero positions $\mathbf{r}_0^{(1)}$ and $\mathbf{r}_0^{(2)}$. This is numerically verified and shown in {\color{red}Fig.~\ref{fig:s5}(d-f)}.

\section{Transition between FCI states with many body Chern numbers $C_\text{mb}=2/3$ and $C_\text{mb}=1/3$ in the moir\'e system with $p4mm$ symmetry and two flatbands per sublattice at $\nu=4/3$}
As we discussed in the main text, there is a dichotomy between interpretations of the two flat bands per sublattice: (i) a Wannier band with $C=0$, and a Chern band $C=1$, (ii) a single Chern band with $C=1$ folded into two bands in a smaller Brillouin zone. At some filling fraction, the two interpretations give two different types of ground states, and only one of those is the actual ground state of the system. This best exemplified at filling fraction $\nu=4/3$. In this case, the first interpretation would suggest the Wannier band is fully filled, and the Chern band is $1/3$ filled giving an FCI state with many-body Chern number $C_\text{mb} =1/3$. However, the second interpretation would suggest that $\nu=4/3$ really is $\nu_\text{unfolded}=2/3$ in the unfolded larger Brillouin zone, and the resulting ground state is an FCI with many-body Chern number $C_\text{mb}=2/3$. In Fig.~\ref{fig:s2} we show that indeed both type of ground states can occur, and by varying a parameter in the single particle Hamiltonian we can induce a phase transition between the two many-body ground states without closing the single particle gap. All details of the system and numerics are given in Fig.~\ref{fig:s2}.
\begin{figure}[ht]
     \centering
\includegraphics[width=\textwidth]{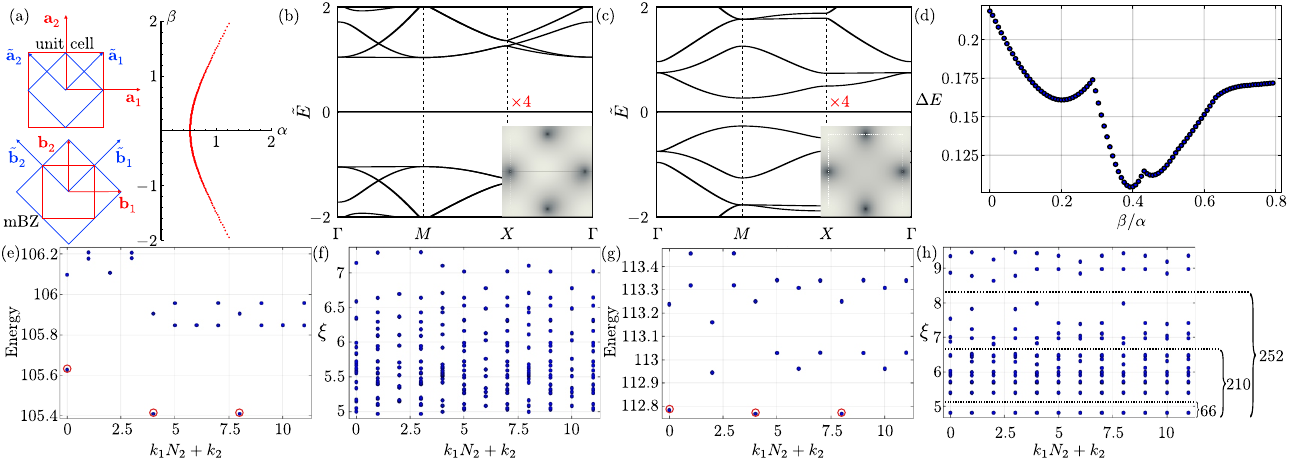}
     \caption{
     Moir\'e system with $p4mm$ symmetry and two flatbands per sublattice, and transition between FCI states with many body Chern numbers $C_\text{mb}=2/3$ and $C_\text{mb}=1/3$  at filling fraction $\nu=4/3$. The single particle Hamiltonian is given in Eq.~\eqref{eq:HSinglep4mm}. (a) shows the moir\'e unit cell and Brillouin zone (mBZ) in red. When $\beta = 0$ in Eq.~\eqref{eq:HSinglep4mm}, the primitive unit cell and corresponding Brillouin zone are shown in blue. The set of parameters $(\alpha,\beta)$, at which four exact flat bands appear, is shown in red line in the $\alpha-\beta$ plane. The band structures for two points ($\alpha=0.5279,\beta=0.0$) and ($\alpha=0.6651,,\beta=0.7183$) on this red line are shown in (b) and (c), respectively. In both (b) and (c), there are four flat bands (two per sublattice) at $E=0$. Since $\beta = 0$ for the plot in (b), the two flat bands per sublattice in (b) can be unfolded into one flat band per sublattice in the bigger blue Brillouin zone if we choose the smaller blue unit cell shown in (a). However, this is not possible in (c) since $\beta\neq 0$. In the bottom right corner of (b) and (c), the corresponding wavefunction $\psi_\Gamma(\mathbf{r})$ is shown. The Dark spots in the wavefunctions are the zeros. There are two zeros per unit cell, which allow for construction of two sublattice-polarized exact flat band wavefunctions as described in Sec.~\ref{sec:FourFlatBand}. The many-body energy spectrum at filling fraction $\nu=4/3$ evaluated via exact diagonalization for ($\alpha=0.5279,\beta=0.0$) and ($\alpha=0.6651,\beta=0.7183$) are shown in (e) and (g), respectively. For exact diagonalization, we take a $(N_1=3)\times(N_2=4)$ system. In both cases, we find three quais degenerate ground states (encircled with red circles). We numerically find the many-body Chern numbers of these quasi-degenerate ground states to be $C_\text{mb}=2/3$ and $C_\text{mb}=1/3$ for (e) and (g), respectively. The corresponding PES for $N_A=2$ (see Sec.~\ref{sec:PES}) are plotted in (f) and (h), respectively. The counting of the low-lying states of PES in (h) matches with a ground state with one fully filled Wannier band and a $1/3$ filled Chern band (see Sec.~\ref{sec:PES} for details). The PES in (f) does not show any gap. (d) shows the many-body gap between the lowest energy excited state and highest energy ground state (among the three quasi-degnerate ground state) as the ratio $\beta/\alpha$ is varied along red line in $\alpha-\beta$ plane in (a). The many-body gap has a global minimum around $\beta/\alpha\approx0.39$. We numerically find that for $\beta/\alpha<0.39$, the many-body Chern number of ground states is $C_\text{mb}=2/3$, whereas for $\beta/\alpha<0.39$, the many-body Chern number of ground states is $C_\text{mb}=1/3$. We conjecture that the many-body gap does not fully close due to finite size effect.
    }
     \label{fig:s2}
\end{figure}
\clearpage
\section{More details of the FCI state with many body Chern number $C_\text{mb}=1/3$ in the moir\'e system with $p6mm$ symmetry and two flatbands per sublattice considered in Fig.~1 of the main text at $\nu=4/3$}

\begin{figure}[ht]
     \centering
\includegraphics[width=\textwidth]{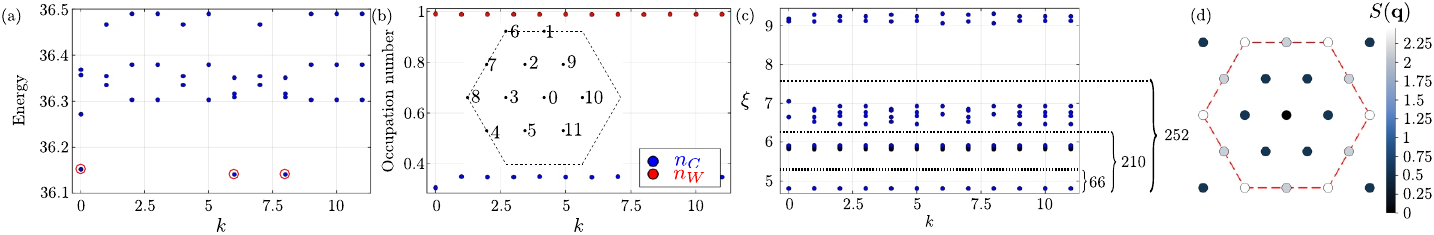}
     \caption{FCI state with many body Chern number $C_\text{mb}=1/3$ in the moir\'e system with $p6mm$ symmetry and two flatbands per sublattice considered in Fig.~1 of the main text at $\nu=4/3$. Many-body energy spectrum (a), ground state occupation number in the Wannier and Chern bands (b), PES for $N_A=2$ (Sec.~\ref{sec:PES}) (c), and the density-density correlation $S(\mathbf{q}) = \langle \rho(\mathbf{q})\rho(-\mathbf{q})\rangle - \langle \rho(\mathbf{q})\rangle\langle \rho(-\mathbf{q})\rangle$ (d) are shown. Three quasi-degenerate ground states encircled by red circles can be seen in (a). (b) shows that the Wannier band is fully filled, whereas the Chern band is $1/3$ filled. (c) The gaps in PES are consistent with PES counting for ground state with a fully filled trivial band and a $1/3$ filled Chern band (see Sec.~\ref{sec:PES} for details). 
    }
     \label{fig:s4}
\end{figure}

\section{FCI state at filling fraction $\nu=5/3$ with many body Chern number $C_\text{mb}=2/3$ in the moir\'e system with $p6mm$ symmetry and two flatbands per sublattice considered in Fig.~1 of the main text}
\begin{figure}[ht]
     \centering
\includegraphics[width=\textwidth]{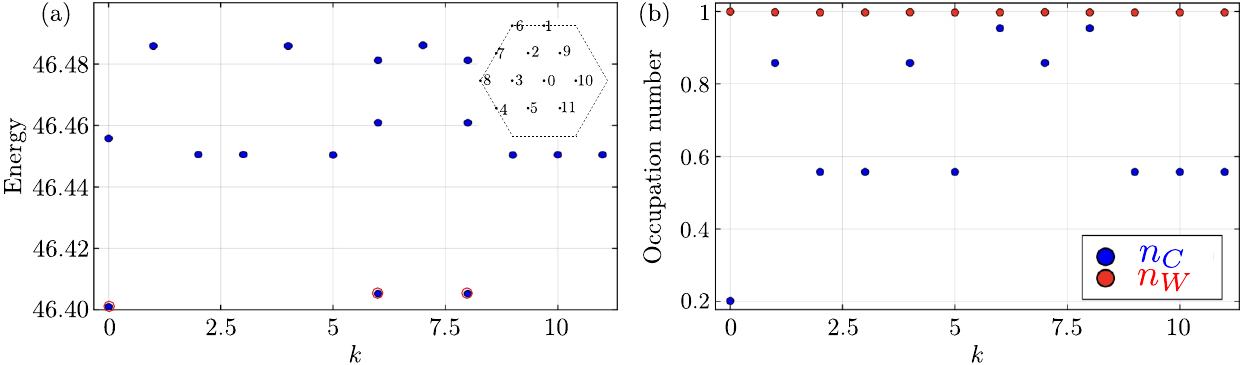}
     \caption{
     Many-body energy spectrum (a) and occupation number in the Chern ($n_C$) and Wannier ($n_W$) bands of the ground-states (b) at filling $\nu=5/3$. The three quasi-degenerate ground states are marked by red circles in (a). In the ground state, the Wannier band is fully filled, whereas the Chern band is $2/3$ filled. We numerically find the many-body Chern number of the ground states $C_\text{mb}=2/3$.
    }
     \label{fig:s7}
\end{figure}

\section{Particle entanglement spectrum}\label{sec:PES}
Information about the nature of ground state can be obtained from particle entanglement spectrum (PES)~\cite{li2008entanglement,sterdyniak2011extracting}. For example, the number of ground state for $\sqrt{3}\times\sqrt{3}$ charge density wave and a $\nu=1/3$ or $\nu=2/3$ FCI state are the same, and they appear at the same center of mass momenta in exact diagonalization. In these cases PES helps in pinpointing the ground state, since it provides correct counting of the number of quasihole excitations. For $d$-fold quasi-degenerate many-body ground states $|\Psi_\text{mb}^{(i)}\rangle$ with $N_e$ electrons, consider the density matrix $\rho = \sum_{i=1}^d\frac{1}{d}|\Psi_\text{mb}^{(i)}\rangle\langle\Psi_\text{mb}^{(i)}|$. Dividing the $N_e$ electrons into two groups with $N_A$ and $N_B$ number of electrons, we trace out the $N_B$ electrons to obtain $\rho_A= \text{tr}_B{\rho}$. We follow Ref.~\cite{wu2024quantum} for the implementation of this procedure. Let the eigenvalues of $\rho_A$ be $\exp(-\xi)$; we plot $\xi$'s vs the center of mass momentum of the $N_A$ particles. We generally expect a gap in PES, and the states of PES below  are exactly the quasi-hole states~\cite{sterdyniak2011extracting}. Below, we give the counting for each state (CDW or FCI) encountered in the main text.

\subsubsection{$\nu=2/3$ CDW state in Fig.~\ref{fig:s1}(g-j)}
In the case of $\sqrt{3}\times\sqrt{3}$ CDW state at $\nu=2/3$ filling of the Wannier band (the Chern band is empty as can be seen from Fig.~\ref{fig:s1}(h)) with $N=12$ single particle states and $N_e=8$ electrons, we find that there is a gap above $18$ states and another gap above $66$ states for $N_A=2$. First we discuss why there is a gap above 18 states. Note that the $N=12$ discretization of the BZ means that the system has 12 moir\'e unit cells. Because CDW enlarges the unit cell to three times the area of the moire unit cell, there are only 4 unit cells in the CDW state. The $N_A=2$ electrons can be at two of these 4 unit cells in $\binom{4}{2} =6$ different ways. However, there are 3 equivalent CDW states. So, the $N_A=2$ electrons can be at either of these three CDWs. Therefore, there are in total $3\times\binom{4}{2} = 18$ different states. This is why there is a gap above 18 states. There is also a gap above 66 states. This can be understood in the following way. If the the two electrons go to two different CDW states, and each electron chooses 1 of the 4 unit cells available in each CDW state, then we have $\binom{3}{2}\binom{4}{1}\binom{4}{1} = 48$ states. This is why there is a gap above $48+18=66$ states.

\subsubsection{FCI state at $\nu=2/3$ with many-body Chern number $C_\text{mb}=1/3$ in Fig.~2 of the main text and Fig.~\ref{fig:s1}(c-f)}
In this article our claim is that the many-body Chern number $C_\text{mb}=1/3$ of this FCI state can be explained by thinking about the two bands (per sublattice) as one $C=1$ Chern band folded into a smaller Brillouin zone of half the area such that filling fraction in the unfolded Brillouin zone the filling fraction is $\nu_\text{unfolded} =1/3$. The gap above 228 states in the PES of this state for $N_A=2$ and $N_e=8$ as shown in Fig.~\ref{fig:s1}(e) is consistent with claim. To see this, since the BZ has 12 momentum points, in the unfolded larger BZ, the number of momentum points (or in other words the number of single particle states in the Chern band in the unfolded BZ) is $2\times 12 =24$. Now, according to~\cite{regnault2011fractional}, the PES of a $\nu=1/3$ state with $N_A=2$ and $N=24$ single particles states is $N\frac{(N-2N_A-1)!}{(N_A!)((N-3N_A)!)} = 24 \frac{(19)!}{(2!)((18)!)}=228$, which is exactly what we find below gap.

\subsubsection{FCI state at $\nu=4/3$ with many-body Chern number $C_\text{mb}=1/3$ in Fig.~2 of the main text and Figs.~\ref{fig:s2}(g-h)}
In this case we saw in Fig.~2 of the main text that the Wannier band is fully filled and the Chern band is one-third filled. Hence, the $N_A=2$ electrons can both go to the Wannier band (with $N=12$ single particle states), in which case the number of PES states would be $\binom{12}{2} = 66$. Alternatively, one can go to Wannier band, and the other can go to the Chern band, in which case the number of states is $\binom{12}{1}\binom{12}{1} =144$. Otherwise, both can go to the Chern band, in which case the number of states would be $N\frac{(N-2N_A-1)!}{(N_A!)((N-3N_A)!)} = 12 \frac{(7)!}{(2!)((6)!)}=42$ as per Ref.~\cite{regnault2011fractional}. This is why there are gaps above 66, $66+144=210$, and $66+144+42 = 252$ states in the PES in {\color{red}Figs.~\ref{fig:s2}(h) and~\ref{fig:s4}(c)}.

\subsubsection{FCI state at $\nu=4/3$ with many-body Chern number $C_\text{mb}=2/3$ in Fig.~\ref{fig:s2}(e-f)}
In this case, when $\beta = 0$, the chosen unit cell is twice as large as the primitive moir\'e unit cell (as shown in {\color{red}Fig.~\ref{fig:s2}(a)}). Hence the chosen Brillouin zone is half the Brillouin zone corresponding to the primitive unit cell. Hence, two flat bands per sublattice are folded version of one $C=1$ Chern band. Hence, $\nu=4/3$ filling of corresponding to the smaller Brillouin zone is actually $\nu_\text{unfolded} =2/3$ filling of the unfolded single Chern band. No simple counting exists for PES for $\nu_\text{unfolded}=2/3$ when we partition the electron (PES counting exists if we partition the minority species, i.e. holes). However, we point out that PES in Fig.~\ref{fig:s2}(f) is very different from the PES in Fig.~\ref{fig:s2}(h) in that the former does not have any gaps. 

\section{$\det(n_{ij}(\mathbf{k}))\neq 0$ in Fig. A1 implies the FCI at $\nu=2/3$ filling cannot be single-band-like in any basis}

To see why $\det(n_{ij}(\mathbf{k}))\neq 0$ means the state cannot be single-band-like, consider the following. Let there be a basis where the FCI state is single-band-like. By this we mean that we can do a unitary transformation of the original basis consisting of a Wannier band and Chern band with creation operators $c^\dagger_C(\mathbf{k})$ and $c^\dagger_W(\mathbf{k})$ to a new basis with creation operators  $c^\dagger_{n}(\mathbf{k})$ and $c^\dagger_{n'}(\mathbf{k})$ at every momentum $\mathbf{k}$ such that after the transformation the FCI state is only supported in the band whose creation operators are $c^\dagger_{n}(\mathbf{k})$. We can write these new creation operators $c^\dagger_{n}(\mathbf{k})$ and $c^\dagger_{n'}(\mathbf{k})$ of the new basis at each momentum $\mathbf{k}$ as $c^\dagger_{n}(\mathbf{k}) = u_\mathbf{k} c^\dagger_{W}(\mathbf{k})+v_\mathbf{k} c^\dagger_{C}(\mathbf{k})$ and $c^\dagger_{n'}(\mathbf{k}) = v_\mathbf{k}^* c^\dagger_{W}(\mathbf{k})-u_\mathbf{k}^* c^\dagger_{C}(\mathbf{k})$. 
Since the FCI ground state is single-band-like and only supported in the band whose creation operators are $c^\dagger_{n}(\mathbf{k})$, this state can be written as
\begin{equation}
    |\Psi_\text{GS}\rangle = \sum_{I} a_I \prod_{J=1}^{N_e} c_{n}^\dagger(\mathbf{k}_J^{(I)})|0\rangle,
\end{equation}
where $I$ goes over all the Slater determinant states with $N_e$ number of electrons in band $n$. We can evaluate $n_{ij}(\mathbf{k})$:
\begin{equation}
    \begin{split}
        n_{CC}(\mathbf{k}) &= \langle \Psi_\text{GS}| c^\dagger_C(\mathbf{k})c_C(\mathbf{k})|\Psi_\text{GS}\rangle\\
        &=\langle \Psi_\text{GS}| (v_\mathbf{k}^*c^\dagger_n(\mathbf{k})-u_\mathbf{k}c^\dagger_{n'}(\mathbf{k}))(v_\mathbf{k}c_n(\mathbf{k})-u_\mathbf{k}^*c_{n'}(\mathbf{k}))|\Psi_\text{GS}\rangle\\
        &=|v_\mathbf{k}|^2 \langle \Psi_\text{GS}| c^\dagger_n(\mathbf{k})c_n(\mathbf{k})|\Psi_\text{GS}\rangle,\\
        n_{WW}(\mathbf{k}) &= \langle \Psi_\text{GS}| c^\dagger_W(\mathbf{k})c_W(\mathbf{k})|\Psi_\text{GS}\rangle\\
        &=\langle \Psi_\text{GS}| (u_\mathbf{k}^*c^\dagger_n(\mathbf{k})+v_\mathbf{k}c^\dagger_{n'}(\mathbf{k}))(u_\mathbf{k}c_n(\mathbf{k})+v_\mathbf{k}^*c_{n'}(\mathbf{k}))|\Psi_\text{GS}\rangle\\
        &=|u_\mathbf{k}|^2 \langle \Psi_\text{GS}| c^\dagger_n(\mathbf{k})c_n(\mathbf{k})|\Psi_\text{GS}\rangle,\\
        n_{CW}(\mathbf{k}) &= \langle \Psi_\text{GS}| c^\dagger_C(\mathbf{k})c_W(\mathbf{k})|\Psi_\text{GS}\rangle\\
        &=\langle \Psi_\text{GS}| (v_\mathbf{k}^*c^\dagger_n(\mathbf{k})-u_\mathbf{k}c^\dagger_{n'}(\mathbf{k}))(u_\mathbf{k}c_n(\mathbf{k})+v_\mathbf{k}^*c_{n'}(\mathbf{k}))|\Psi_\text{GS}\rangle\\
        &=u_\mathbf{k}v_\mathbf{k}^* \langle \Psi_\text{GS}| c^\dagger_n(\mathbf{k})c_n(\mathbf{k})|\Psi_\text{GS}\rangle,
    \end{split}
\end{equation}
where we used the fact that $c_{n'}(\mathbf{k})|\Psi_\text{GS}\rangle = 0$. Hence the occupation number matrix at momentum $\mathbf{k}$ is of the form
\begin{equation}
    n_{ij}(\mathbf{k}) = \langle \Psi_\text{GS}| c^\dagger_n(\mathbf{k})c_n(\mathbf{k})|\Psi_\text{GS}\rangle \begin{pmatrix}
        |v_\mathbf{k}|^2 & u_\mathbf{k}v_\mathbf{k}^*\\ u_\mathbf{k}^*v_\mathbf{k} & |u_\mathbf{k}|^2
    \end{pmatrix},
\end{equation}
and its determinant is $\det( n_{ij}(\mathbf{k})) = (\langle \Psi_\text{GS}| c^\dagger_n(\mathbf{k})c_n(\mathbf{k})|\Psi_\text{GS}\rangle)^2 (|u_\mathbf{k}|^2|v_\mathbf{k}|^2-u_\mathbf{k}v_\mathbf{k}^*u_\mathbf{k}^*v_\mathbf{k}) = 0$.

Hence, above we proved that \textit{if the ground state is single-band-like in any basis ($u_\mathbf{k}$ and $v_\mathbf{k}$ are arbitrary), then the determinant of $n_{ij}(\mathbf{k})$ must be zero. We find that $\det(n_{ij}(\mathbf{k}))$ is nonzero for $\nu=2/3$ FCI state at all $\mathbf{k}$ (as shown in Fig. A1). Hence, this FCI cannot be single-band-like.}

\section{Possible experimental realization}

The main ingredients for the physics discussed in this paper are (i) Effective $2n\pi, n>1$ magnetic flux per moir\'e unit cell, which gives rise to $n$-flatbands per spin/valley/sublattice. In a previous paper~\cite{sarkar2023symmetry}, we enumerated the symmetries and number of tuning parameters required for the realization of multiple flatbands per spin/valley/sublattice. Once these requirements are met, multiple flat bands per spin/valley/sublattice can be achieved by tuning the parameters. (ii) short range repulsive interaction which generally favors FCI state.

Both twisted bilayer MoTe$_2$ and twisted bilayer graphene have $2\pi$ flux per moir\'e unit cell~\cite{ledwith2020fractional,morales2024magic}, that is why they have single band per spin/valley/sublattice. On the other hand, rhombohedral graphene at the single particle level does not have isolated flat bands, and there is ongoing debate on the origin of FCI in this system~\cite{yu2025moire}. 

One interesting platform where this physics can be realized is twisted bilayer graphene with spacially alternating magnetic field as described in~\cite{le2022double,sarkar2023symmetry}. This system has four flat bands (two per sublattice) per valley per spin, and in the chiral limit  the high symmetry momentum wavefunction $\psi_\Gamma(\mathbf{r})$ has two zeros~\cite{sarkar2023symmetry} at the corners of the moir\'e unit cell indicating that the magnetic flux per moir\'e unit cell is $4\pi$.

Another pathway is to use 2D materials with quadratic band crossing at the Fermi level under periodic strain field. Several such materials have been predicted: PO~\cite{zhu2016blue}, Mg$_2$C~\cite{wang2018monolayer}, Pd$_3$P$_2$S$_8$~\cite{park2020kagome}, Cu$_2$N~\cite{hu2023realization}, metal organic framework material Cr$_3$(HAB)$_2$~\cite{feng2024quadratic} and at $K$ point such as bernal bilayer graphene~\cite{jung2014accurate,wan2023nearly}. Furthermore, experimental techniques for applying periodic strain field have also been rapidly evolving;
such as using nanopillars/nanospheres~\cite{jiang2017visualizing,zhang2019magnetotransport}, buckling~\cite{mao2020evidence}, prestretched substrate with 2D material undergoing wrinkling upon release of stretch~\cite{cho2021highly}, process-induced strain~\cite{zhang2024patternable,zhang2024enhancing}, etc  (see~\cite{kim2023strain} for a review of strain engineering in 2D materials), some of which can indeed give rise to nearly periodic rotationally symmetric strain field.

\end{document}